\documentclass[12pt, draftclsnofoot, onecolumn]{IEEEtran}

\usepackage[T1]{fontenc}

\usepackage{cite}
\usepackage{graphicx}
\usepackage{amsmath}
\usepackage{amsthm,bm}
\usepackage{algorithmic}
\usepackage{algorithm}
\usepackage{array}
\usepackage{stfloats}
\usepackage{url}
\usepackage[caption=false,font=footnotesize,labelfont=sf,textfont=sf]{subfig}
\usepackage{amsfonts}

\graphicspath{{./figures/}}

\begin{document}
\title{Clustering Strategies for Multicast Precoding\\ in Multi-Beam Satellite Systems}


\author{Alessandro~Guidotti,~\IEEEmembership{Member,~IEEE,}\\
        Alessandro~Vanelli-Coralli,~\IEEEmembership{Senior Member,~IEEE}
        % <-this % stops a space
\thanks{A. Guidotti and A. Vanelli-Coralli are with the Department of Electrical, Electronic, and Information Engineering, University of Bologna, 40134, Bologna, Italy (email:\{a.guidotti,alessandro.vanelli\}@unibo.it).}% <-this % stops a space
% <-this % stops a space
%\thanks{Manuscript submitted for review on June 28, 2017.}
}

%\markboth{IEEE TRANSACTIONS ON WIRELESS COMMUNICATIONS}{Guidotti \MakeLowercase{\textit{et al.}}: User Clustering for Multicast Precoding in Multi-Beam Satellite Systems}

\maketitle

%\IEEEtitleabstractindextext{%
\begin{abstract}
Next generation multi-beam SatCom architectures will heavily exploit full frequency reuse schemes along with interference management techniques, \emph{e.g.}, precoding or multiuser detection, to drastically increase the system throughput. In this framework, we address the problem of the user selection for multicast precoding by formulating it as a clustering problem. By introducing a novel mathematical framework, we design fixed-/variable-size clustering algorithms that group users into simultaneously precoded and served clusters while maximising the system throughput. Numerical simulations are used to validate the proposed algorithms and to identify the main system-level trade-offs. 
\end{abstract}

%}

\IEEEpeerreviewmaketitle

\IEEEdisplaynontitleabstractindextext
% compsoc or transmag under a non-conference mode

% \fi
%\IEEEpeerreviewmaketitle

\section{Introduction}
\IEEEPARstart{T}{his} work addresses the selection of the users to be multiplexed together in multicast precoding for multi-beam High Throughput Satellite (HTS) systems with full frequency reuse. In order to provide large throughput to the on-ground users, Satellite Communication (SatCom) systems evolved towards advanced multi-beam deployments with a very large number of beams covering a single region, \emph{e.g.}, more than $100$ beams for Europe, \cite{Intro5}. The increased number of beams for a fixed on-ground coverage brings into the SatCom context frequency reuse concepts similar to those already driving the design of cellular systems. State-of-the-Art (SoA) HTS systems typically split the available system bandwidth into two frequency bands with two orthogonal polarisations, leading to a 4-colour frequency reuse scheme. In order to provide terabit connectivity, different architectures and technologies are considered aimed at increasing the system throughput by either augmenting the available spectrum or the system level spectral efficiency. Notably, current physical layer (PHY) technologies already bring the link spectral efficiency close to the theoretical limits, thus moving the emphasis on the design of system level techniques. In this framework, during the last years, research focused on increasing the available spectrum bandwidth, either by adding spectrum chunks, as in flexible spectrum usage paradigms based on spectrum sharing through Cognitive Radio techniques, \cite{Intro1,Intro2,Intro7}, or by decreasing the frequency reuse factor down to the \emph{full frequency reuse} case. The latter case, which is the focus of this paper, calls for the design of interference management techniques to counterbalance the significantly increased co-channel interference. In multi-beam SatCom systems, severe co-channel interference environments arise because of the side-lobes of the beams radiation patterns in the iso-frequency allocation paradigm and the interference mitigation techniques implemented to limit its disruptive effect mainly include transmitter side techniques, \emph{e.g.}, precoding, \cite{Prec1,Prec2,ESAMIMO}, and receiver side techniques, \emph{e.g.}, Multi-User Detection (MUD), \cite{Intro8}.

In particular, the success of multi-user Multiple-Input Multiple-Output (MU-MIMO) techniques in terrestrial communications, together with the introduction of the super-frame structure in the DVB-S2X standard, \cite{DVBS2X}, led the Satellite Communications community to assess and implement precoding techniques in multi-beam HTS systems. Building on the early works in \cite{Prec1,Prec2,Prec3}, which paved the way for the application of MU-MIMO in HTS systems, the SatCom community has been intensifying the work on precoding-based satellite systems, as demonstrated by the significant contributions brought to this topic in \cite{Prec4,Prec5,Prec6,Prec7,Prec8,Prec8_1,Prec10,Prec11}. These works showed that significant gains can be obtained in the overall system throughput by means of both linear and non-linear precoding techniques over the multi-beam fixed satellite channel. Since linear precoding techniques provide already significant throughput gains with a limited complexity, they might be preferred with respect to more complex techniques, \emph{i.e.}, non-linear precoding, for which the increase in the system complexity is not justified by an equivalent performance benefit, \cite{Intro9}. In this context, multicast precoding techniques have been recently proposed, \cite{Prec12,Prec13,Prec14,CTTCprec,Taricco}: differently from the unicast case, in multicast precoding, multiple users are multiplexed into a single codeword that is then precoded according to an equivalent channel obtained as the average of the users' channel coefficients vectors. Since the average channel coefficients exploited to compute the precoding vector depend on the selected users and the modulation and coding scheme (ModCod) used for the multiplexed users depends on the worst-case user, \emph{i.e.}, the user with lowest Signal-to-Interference-plus-Noise Ratio (SINR), so as to guarantee that all users can decode their information, user selection and grouping in the same codeword directly and deeply impacts the overall system performance in terms of achievable throughput, \cite{Taricco,Prec14,Prec8}. In this paper, we propose a novel framework based on \emph{clustering theory} concepts for the users grouping in multicast precoding. This new framework allows us to identify and highlight the pivotal aspects to be taken into account by the designer of a multi-beam HTS system with multicast precoding.

\subsection{Previous works}
The first implementations of the precoding concept to a Satellite Communication context is provided in \cite{Prec1,Prec2}. In these early works, the authors moved from the linear MU-MIMO techniques already substantiated in terrestrial cellular systems and proposed both Zero-Forcing (ZF) and Minimum Mean Square Error (MMSE) algorithms to enhance the throughput on the Forward Link (FL) of a multi-beam satellite system. These works also provided some considerations on the main implementation challenges that impact the effectiveness of precoding in SatCom systems, together with \cite{Prec3}, in which the impact of partial Channel State Information (CSI) at the transmitter (CSIT) has been addressed. In \cite{Prec11}, the implementation of precoding to the DVB-S2X standard is discussed based on the outcome of several European Space Agency (ESA) R\&D activities. In particular, practical challenges related to the implementation of precoding to HTS systems are discussed, like framing issues, imperfect channel estimation, outdated phase estimates, and multiple gateways. An extensive analysis of practical impairments for the implementation of precoding in DVB-S2X systems is also provided in \cite{Prec6}. In \cite{Prec5}, the authors provide a review of several precoding techniques and propose an optimisation of the linear precoding design. In particular, general linear and non-linear power constraints are addressed by means of an iterative algorithm that optimises the precoding vectors and the power allocations in an alternating fashion. An extensive review of precoding techniques for multi-beam systems is also provided in \cite{Prec7}. The authors of \cite{Prec4} implemented the Tomlinson-Harashima precoding (THP), a non-linear precoding technique based on modulo operations over the constellation symbols, also taking into account the beam gain. In \cite{Prec10}, the authors assess the performance of linear beamforming in terms of satisfying specific traffic demands by including generic linear constraints the transmit covariance matrix. 

The first work providing the design of multicast precoding for satellite systems was based on a regularised channel inversion, \cite{Taricco}. In this work, the author proposed to consider all the users to be served together as a single user with an equivalent channel matrix equal to the average of the single channel matrices. In this work, a geographical user grouping is also proposed in which users to be precoded together are chosen based on their geographical position, under the assumption that the same number of users are precoded together across the different beams. The computation of the precoding matrix is based on the pragmatic approach proposed in \cite{CTTCprec}, in which the authors jointly design the linear precoding and ground-based beamforming at the gateway. An alternative solution to the computation of the precoding matrix is provided in \cite{Prec8_1}, in which the authors implement a technique based on block Singular Value Decomposition (SVD). In this case, the precoding matrix is built row-wise and the performance is actually improved, although with a significant increase in the system computational cost. The authors of \cite{Prec8} discuss on the implementation of linear precoding techniques to multi-beam broadband fixed satellite communications by also also introducing some preliminary consideration on the issues related to users grouping. In \cite{Prec12}, the optimisation problem of multicast precoding with per-antenna power constraints is addressed, for which the authors propose an optimal, although computationally costly, solution. In \cite{Prec13}, the authors focus on framing multiple users per transmission and on the per-antenna transmit power limitations and propose a solution for frame-based precoding based on the principles of physical layer multicasting to multiple co-channel groups under per-antenna constraints. Finally, in \cite{Prec14}, a two stage linear precoding is proposed to lower the complexity in the ground segment, under the presence of non-ideal CSI as well. In addition, some aspects related to user grouping are also discussed, in particular referring to a grouping based on the channel coefficients. The problem of multicasting precoding has recently started to be addressed in terrestrial networks as well. In \cite{PrecTerr1}, the authors focus on the transmission of the same message from a distributed set of transmitting antennas to several targeted beams, while implementing nullforming towards the remaining beams. The proposed distributed algorithm exploits a local knowledge at each transmitting antenna, \emph{i.e.}, each transmitting antenna only has knowledge of the channel coefficients between itself and the users. Joint beamforming has been addressed in \cite{PrecTerr2}, in which the authors propose a design among multiple transmitting base stations for Long Term Evolution (LTE) Coordinated MultiPoint (CoMP) transmissions, based on different optimisation algorithms to solve the max-min fair problem for the Multimedia Broadcast Multicast Service (MBMS). The design of higher rank transmissions to increase the spectrum efficiency and the design of robust beamforming to alleviate performance degradation caused by imperfect channel state information are also addressed. 

In the literature, aspects related to the choice of the users to be grouped together within the same codeword were mainly addressed in \cite{Taricco,Prec14,UniLu_New}. In particular, the proposed algorithms take as an input the metric to be used to group the users, \emph{i.e.}, either the locations in the Euclidean space, \cite{Taricco}, or the channel vectors, \cite{Prec14,UniLu_New}, and then provide a group for each beam with a predefined cardinality $K$ assumed to be the same across all beams. In \cite{Taricco,Prec14}, the proposed grouping assumes that, at each time frame, there is a new randomly selected user in each beam coverage independently of whether it has already been served or not. This is a strong assumption at system level in terms of overall fairness and, in addition, it also does not properly manage the potential \emph{outliers}, \emph{i.e.}, users that, based on the considered metric, are located far away from the others. As a matter of facts, if the randomly chosen reference user is an outlier, the system performance will be deeply degraded due to the limited similarity in terms of channel coefficients among the group members. The fairness issue has been circumvented in \cite{UniLu_New}, in which the authors, focusing on the channel coefficients similarities, implement a similar user selection algorithm, but having memory of those already served in the previous time frames, \emph{i.e.}, at each time frame, the reference user is randomly chosen among those that have not been served yet. Although the proposed solution is a valuable advance in the analysis of user grouping algorithm, it still does not take into account system \emph{outliers}. When the system is grouping the last few and yet unserved users, in fact, there might be situations in which they are located far from each other, thus leading back to the outliers challenge highlighted above.

\subsection{Contributions and Paper Organisation}

Moving from the previous works and the multicast precoding implementations to SatCom in which the user grouping was not the pivotal aspect, in this paper we provide the following contributions:
\begin{itemize}
    \item Definition of a mathematical framework for the design of multicast precoding systems based on clustering theory and algorithms. In this context, details on the initial selection of the users so as to avoid the sensitivity to system outliers are provided.
    \item Design of two clustering algorithms, one with fixed-size and one with variable-size clusters, aimed at reducing the impact of system outliers through proper clusters initialisation.
    \item Discussion of an extensive and thorough numerical simulation campaign aimed at identifying the main limiting factors and benefits in implementing proper clustering algorithms for HTS based on multicast precoding, both in terms of average spectral efficiency and of clusters' homogeneity in the considered similarity space.
    \item Comparison of the performance of the proposed algorithm with those provided in the literature assuming two similarity metrics: i) the Euclidean distance; and ii) the distance in the multi-dimensional space of user channel coefficients.
\end{itemize}
To the best of the authors' knowledge, this is the first paper in which the users' grouping in multicast precoding is tackled in a mathematical framework based on clustering theory and the algorithms performance is assessed not only in terms of average rate but also from the clusters' homogeneity perspective. The remainder of this paper is organised as follows. In Section II, we introduce the system model in terms of system architecture, multicast precoding algorithm, and problem statement. In Section III, we formulate the users grouping problem of multicast precoding in terms of a clustering problem and propose two new clustering algorithms. In addition, the algorithms found in the literature used as benchmark for the proposed solutions are also detailed in the novel framework. In Section IV, we provide the numerical results of the extensive simulation campaign in which the average rate per beam is obtained as a function of the number of users precoded in the same frame. In addition, the algorithms performance is also assessed in terms of clusters' homogeneity. Finally, Section V concludes this paper.

\paragraph*{Notation}
Throughout this paper, and if not otherwise specified, the following notation is used: bold face lower case and bold face upper case characters denote column vectors and matrices, respectively. ${(\cdot)}^T$ denotes the matrix transposition operator. ${(\cdot)}^H$ denotes the matrix conjugate transposition operator. $\parallel\cdot\parallel$ denotes the Euclidean norm. $\mathbb{E}\left\{\cdot\right\}$ denotes the expectation operator. $\mathrm{diag}(\cdot)$ denotes a diagonal matrix. $\mathbf{I}_N$ denotes the $N\times N$ identity matrix. $\mathcal{N}_c\left(\mu,\sigma^2\right)$ denotes the circularly-symmetric complex Gaussian distribution with mean $\mu$ and variance $\sigma^2$. $\left|\mathcal{A}\right|$ denotes the cardinality of set $\mathcal{A}$ and $\mathrm{tr}\left(\mathbb{A}\right)$ denotes the trace of matrix $\mathbb{A}$.

\section{System Model and Problem Statement}
\label{sec:SystemModel}
\subsection{Architecture and multicast precoding}
We consider a Geostationary Earth Orbit (GEO) HTS system operating with full frequency reuse to provide broadband connectivity by means of multiple beams. In terms of system architecture and operations, the following assumptions hold throughout the paper, if not otherwise specified: i) the satellite payload is assumed to be transparent and equipped with $N_B$ antennas, generating $N_B$ on-ground beams; ii) linear precoding is implemented on the Forward Link (FL); and iii) a single gateway (GW) manages the Channel State Information (CSI), assumed to be ideal, obtained from the Return Link (RL) in order to compute the precoding weights. We further assume that Time Division Multiple Access (TDMA) is implemented to serve the users in the $N_B$ beams in each time frame, as it is for the majority of the open SatCom air interfaces, as DVB-S2 and DVB-S2X, \cite{DVBS2,DVBS2X}.
%Thus, during each time frame, the GW relies on the CSI previously obtained from the RL and simultaneously serves $N_B$ users by implementing linear precoding algorithms.

\begin{figure}[!t]
\centering
\includegraphics[width=0.7\columnwidth]{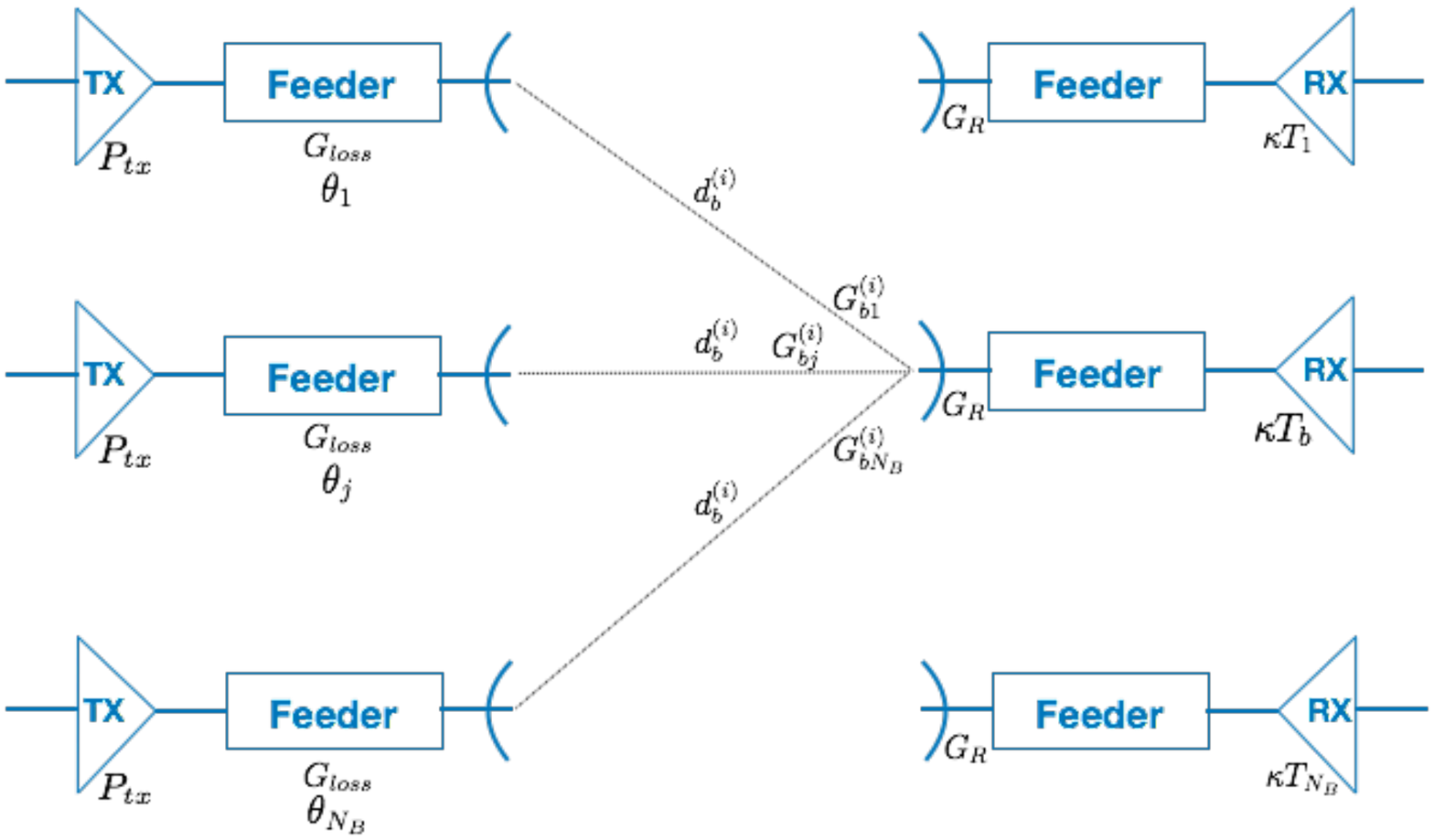}
\caption{Channel elements for the generic $i$-th user in the $b$-th beam.}
\label{fig:Channel}
\end{figure}

\subsubsection{User deployment and channel model}
Considering the $b$-th beam, we assume $N_U^{(b)}$ uniformly distributed users, being $N_U^{(b)}=\left[\rho_b A_b\right]$, where $A_b$ is the $b$-th beam area in [km$^2$] and $\rho_b$ the user density in [users/km$^2$], which in the following is assumed to be equal for each beam, \emph{i.e.}, $\rho_b=\rho=\mathrm{const},\forall b$. The uniformly distributed users are assumed to be deployed at fixed locations, as shown in Fig.~\ref{fig:Beam1_DeployedUsers}. We define by $\mathbf{h}_b^{(i)} = \left(h_{b,1}^{(i)},\ldots,h_{b,N_B}^{(i)} \right)$ the vector of channel coefficients between the generic $i$-th user in the $b$-th beam and the $N_B$ transmitting antennas, where the single elements are defined as follows and shown in Fig.~\ref{fig:Channel}: 
\begin{equation}
\label{eq:ChannelGain}
    h_{bj}^{(i)} = \frac{\sqrt{G_RG_{loss}G_{bj}^{(i)}}}{4\pi \frac{d_b^{(i)}}{\lambda}\sqrt{P_{Z,b}}}e^{-\jmath\frac{2\pi}{\lambda}d_b^{(i)}}e^{-\jmath \vartheta_b},\ j=1,\ldots,N_B
\end{equation}
where $G_R$ is the receiver antenna gain, $G_{loss}$ models the overall antenna losses, $G_{bj}^{(i)}$ is the multi-beam antenna gain between the $j$-th antenna feed and the $b$-th receiving beam, $d_b^{(i)}$ is the distance between the GEO satellite and the considered $i$-th user in the $b$-th beam, $\lambda$ the carrier wavelength, $\vartheta_b\sim\mathcal{U}\left[0,2\pi\right)$ is the random phase offset that depends on the transmitting antenna only. In addition, $P_{Z,b}=\kappa T_b B_w$ is the noise power at the $b$-th receiving antenna, in which $\kappa$ is the Boltzmann constant, $T_b$ the clear-sky noise temperature, and $B_w$ the user's bandwidth. Based on these assumptions, the signal received by the $i$-th user in the $b$-th beam is given by:
\begin{equation}
\label{eq:RX_signal}
    y_b^{(i)} = \sqrt{p_b^{(i)}}\mathbf{h}_b^{(i)}\mathbf{x} + z_b^{(i)}, \ i=1,\ldots,N_U^{(b)}
\end{equation}
where: i) $\mathbf{x}$ is the $N_B\times 1$ vector of complex transmitted symbols; ii) $\sqrt{p_b^{(i)}}$ is the power scaling factor, \emph{i.e.}, the power allocated to the $i$-th user in the $b$-th beam. It shall be noted that this value is inherently different from the power emitted by each antenna, which is assigned by the precoder; and iii) $z_b^{(i)}$ is a complex circularly-symmetric independent and identically distributed (i.i.d.) Gaussian random variable with zero-mean and unit variance, since the noise term is included in the channel coefficients. In the following, we assume that $\sqrt{p_b^{(i)}}=\sqrt{p}=\sqrt{P_{TX}}$, $\forall b,i$.

\subsubsection{Unicast precoding}
In traditional unicast precoding, $N_B$ users are served by the $N_B$ transmit antennas and the overall channel can be modeled as a $N_B\times N_B$ MIMO matrix $\widetilde{\mathbf{H}}$ representing the estimated channel. This matrix is built from the single users' channel vectors $\mathbf{h}_b^{(i)}$ as $\widetilde{\mathbf{H}} = {\left({\left(\mathbf{h}_1\right)}^T,\ldots,{\left(\mathbf{h}_{N_B}\right)}^T\right)}^T$, where we dropped the user index $i$ for the sake of clarity. The estimated channel matrix is then fed to a linear precoding algorithm, which in the following is assumed to be a Minimum Mean Square Error (MMSE) precoder:
\begin{equation}
\label{eq:MMSE_PREC}
    \mathbf{W} = {\left(  \widetilde{\mathbf{H}}^H\widetilde{\mathbf{H}} + \mathrm{diag}\left(\mathbf{\bm{\alpha}}\right)\mathbf{I}_{N_B} \right)}^{-1}\widetilde{\mathbf{H}}^{H}
\end{equation}
where $\mathrm{diag}\left(\bm{\mathbf{\alpha}}\right)$ is a diagonal matrix of regularisation factors, with $\alpha_b = P_{Z,b}/p = P_{Z,b}/P_{TX}$, with $b=1,\ldots,N_B$. It shall be noted that the precoding matrix $\mathbf{W}$ is changed at each time frame, since the users to be served, and the corresponding channel vectors, are different and it modifies the signal transmitted from the satellite antennas so as to maximise the system sum-rate:
\begin{equation}
    \mathbf{y} = \widetilde{\mathbf{H}}\widetilde{\mathbf{x}} + \mathbf{z} = \widetilde{\mathbf{H}}\mathbf{W}\mathbf{P}\mathbf{x} + \mathbf{z}
\end{equation}
where $\widetilde{\mathbf{x}}=\mathbf{W}\mathbf{P}\mathbf{x}$ is the $N_B\times 1$ vector of precoded transmitted symbols and $\mathbf{P}$ is the diagonal matrix of allocated power levels $\mathbf{P}=\mathrm{diag}\left(\sqrt{p},\ldots,\sqrt{p}\right)=\sqrt{P_{TX}}\mathbf{I}_{N_B}$.
%
%which can then be modelled as a Multiple Input Multiple Output (MIMO) channel described by a $N_B\times N_B$ matrix

\subsubsection{SatCom multicasting and multicast precoding}
In SatCom systems, the information to be sent to different users is multiplexed into a single codeword in each time frame and, in addition, their information bits are also interleaved together, making traditional symbol-level unicast precoding solutions unfeasible. To circumvent this issue, multicast precoding has been proposed, in which the same precoding matrix is applied to all of the symbols in the same codeword, \emph{i.e.}, the precoding matrix $\mathbf{W}$ is constant over a time frame, and multiple users are properly selected and multiplexed in the same codeword. In this context, there are two main challenges to be addressed, differently from unicast precoding: i) how to select the users to be multiplexed together in each beam; and ii) how to process the users' channel vectors in each beam. While the former addressed in this paper and it is extensively discussed in the following sections, we now focus on the latter aspect. In particular, the MMSE precoding matrix in (\ref{eq:MMSE_PREC}) is computed based on an input $N_B\times N_B$ estimated channel matrix $\widetilde{\mathbf{H}}$ that is built based on the users' channel vectors. In multicast precoding, from each beam we have, in general, $K_b>1$ users, each with its corresponding channel vector, which shall thus be elaborated so as to obtain a single representative channel vector $\widetilde{\mathbf{h}}_b$, \emph{i.e.}, one row in the channel matrix. To this aim, in \cite{Taricco}, average precoding was proposed in which a simple arithmetic mean is used to obtain an equivalent estimated channel matrix. In particular, in the generic $b$-th beam, the equivalent channel vector is built as $\widetilde{\mathbf{h}}_b=\left(1/K_b\right)\sum_{i=1}^{K_b}\mathbf{h}_b^{(i)}$, which yields to the average estimated channel matrix $\widetilde{\mathbf{H}} = {\left( {\widetilde{\mathbf{h}}_1}^T,\ldots,{\widetilde{\mathbf{h}}_{N_B}}^T \right)}^T$, which is exploited in (\ref{eq:MMSE_PREC}) to build the precoding matrix. Notably, the more representative the equivalent channel vector $\widetilde{\mathbf{h}}_b$ is, the more adapted to the actual channel conditions of the beam multiplexed users the precoding vector will be.

\subsubsection{Precoding normalisation factors}
Specific power constraints can be taken into account by means of proper normalisation coefficients for the precoding matrix. In the following, we consider both Per Antenna Constraint (PAC) and Sum Power Constraint (SPC) MMSE precoders, in which, respectively: i) the rows are normalised to have unit-norm, \emph{i.e.}, each row is normalised to $\sqrt{\sum_{b=1}^{N_B} {\left|\widetilde{h}_{b,i}\right|}^2}$, $i=1,\ldots,N_B$; and ii) the whole precoding matrix is normalised to $\sqrt{N_B/\mathrm{tr}\left(\mathbf{W}\mathbf{W}^H\right)}$.

\subsection{Problem Statement}
\label{sec:ProblemStatement}
As highlighted above, while in (\ref{eq:MMSE_PREC}) the representative matrix $\widetilde{\mathbf{H}}$ is used to compute the precoding matrix, the signals transmitted from the $N_B$ antennas towards the $K_b$ users in each beam will still experience their actual channels, \emph{i.e.}, the signal received by the $i$-th user in the $b$-th beam is $y_b^{(i)} = \mathbf{h}_b^{(i)}\mathbf{w}_bx_b + \sum_{\ell\neq b} \mathbf{h}_b^{(i)}\mathbf{w}_{\ell}x_{\ell} + z_b^{(i)}$,
where $\mathbf{w}_b$ denotes the $b$-th column of the precoding matrix $\mathbf{W}$ and we highlighted the intended and interfering terms. Assuming Gaussian inputs with unit variance, this yields to the following Signal-to-Interference plus Noise-Ratio (SINR):
\begin{equation}
\label{eq:SINR_2}
    \gamma_i^{(b)} = \frac{{p\left| \mathbf{h}_b^{(i)}\mathbf{w}_b \right|}^2}{1 + \sum_{\ell\neq b}{p\left| \mathbf{h}_{b}^{(i)}\mathbf{w}_{{\ell}} \right|}^2},\ i=1,\ldots,K_b
\end{equation}
and to the related maximum achievable rate from either the Shannon formula or standardised FEC thresholds, \emph{e.g.}, DVB-S2X. However, since we are considering a user multicasting system, within each users group, the actual serving rate for all of its members is constrained to the one of the user experiencing the lowest SINR, since all of the users shall be granted with the possibility to decode the received information. Thus, the actual serving rate for the group in the $b$-th beam is given by $\widetilde{\gamma}^{(b)}=\min_{i}\left\{\gamma_i^{(b)}\right\}$. It is straightforward to note that the closer the SINRs among the group members, \emph{i.e.}, the more representative the equivalent channel vector is, the better will be the performance, \emph{i.e.}, the lower will be the loss between the potential rate of a user with higher SINR than that of the worst-case user and the actually achieved one. 

The objective of this paper is thus to define proper user selection strategies aimed at multiplexing in the same codeword users that are experiencing the most similar channel conditions.

\section{Clustering Algorithms for Multicast Precoding}
\label{sec:Clustering}
In this section, we propose two new strategies to multiplex users together in order to maximise the overall system spectral efficiency, \emph{i.e.}, by grouping together users that can be considered similar based on a given similarity metric, and compare them with two SoA strategies. This is a \emph{cluster analysis problem}: we must group a set of objects (users) such that objects in the same group (a \emph{cluster}) have a closer similarity measure (based on a specific metric) with respect to those those in other groups (clusters).\\
Let us denote by $\mathcal{U}^{(b)}=\left\{\mathbf{u}^{(b)}_1,\ldots,\mathbf{u}^{(b)}_{N_U^{(b)}}\right\}$ the set of randomly deployed users in the $b$-th beam, where $\mathbf{u}_{i}^{(b)}=\left(u_{i1}^{(b)},\ldots,u_{id}^{(b)}\right)\in\mathbb{R}^{d}$ is the $d$-dimensional feature vector representing the generic $i$-th user in the space defined by the chosen similarity metric. Within each beam, we seek a partition of the users set $\mathcal{U}^{(b)}$ into $N_K^{(b)}$ clusters, $\mathcal{C}^{(b)} = \left\{\mathcal{C}_1^{(b)},\ldots,\mathcal{C}^{(b)}_{N_K^{(b)}}\right\}$, such that: i) a user exclusively belongs to a single cluster, \emph{i.e.}, $\mathcal{C}_i^{(b)}\bigcap\mathcal{C}^{(b)}_j=\emptyset$, $i,j=1,\ldots,N_K^{(b)}$ and $i\neq j$; ii) the union set of all clusters provides the set of users initially deployed in the beam, \emph{i.e.}, $\bigcup_{i=1}^{N_K^{(b)}}\mathcal{C}_i^{(b)}=\mathcal{U}^{(b)}$; and iii) all clusters have at least one user, \emph{i.e.}, $\mathcal{C}_i^{(b)}\neq\emptyset$, $i=1,\ldots,N_K^{(b)}$. Note that the first condition guarantees that each user is served in one time frame only, as long as not all of the users have been served in the beam, while the second one ensures that all of the users within a beam are served. Two different aspects shall be addressed: i) the choice of the clustering algorithm; and ii) the choice of the parameter(s) to be optimised by the chosen algorithm, \emph{i.e.}, the similarity measure defining the $d$-dimensional space of the vectors in $\mathcal{U}^{(b)}$. With respect to the latter, we consider both the bidimensional Euclidean distance between users, \emph{i.e.}, $\mathbf{u}_i^{(b)}$ is the location of the $i$-th user in the $b$-th beam, and the distance in the $2N_B$-dimensional channel coefficient space, \emph{i.e.}, the concatenation of the real and imaginary parts of the channel coefficients vector $\mathbf{u}_i^{(b)} = \left(\Re\left\{\mathbf{h}_b^{(i)}\right\},\Im\left\{\mathbf{h}_b^{(i)}\right\}\right)$. Finally, note that, differently from previous works, the number of clusters is variable since we consider a variable number of users per beam. In particular, in the $b$-th beam we have $N_K^{(b)}=\left\lfloor N_{U}^{(b)}/K_b\right\rfloor$ clusters, where $K_b$ is the beam cluster size and the floor operation is needed so as to guarantee an integer number of clusters. In the following, based on the above mathematical framework, we propose two clustering algorithms, one with a fixed cluster size and one with a variable cluster size, aimed at improving the clustering performance in terms of similarity measure. In addition, two algorithms from the literature are outlined and used as a benchmark for the performance evaluation of the novel clustering algorithms.

\subsection{Benchmark algorithms}
In \cite{Taricco,Prec14,UniLu_New}, the authors proposed fixed size (\emph{i.e.}, $K_b=K$ for all beams and clusters) multiplexing strategies that will be used as benchmark in the following. They are based on the choice of a reference user from those uniformly dropped in each beam and on the computation of the $K-1$ closest users with respect to the chosen similarity metric. In particular, in \cite{Taricco,Prec14} the authors considered the Euclidean distance as similarity metric, while in \cite{UniLu_New} the channel coefficients are taken into account. Moreover, both algorithms assume the same number of users across all beams, \emph{i.e.}, $N_U^{(b)} = N_U, \forall b$, which leads to the same number of clusters for all beams, $N_K^{(b)} = N_K$. In the following, we discuss these algorithm by also extending them to the more general case in which: i) the similarity metric can be the distance in either the bidimensional Euclidean space or in the $2N_B$-dimensional channel coefficient space; and ii) the number of users, and, thus, the number of clusters, is variable across the beams based on the user density.
\subsubsection{UpperBound clustering}
A simple approach to user clustering is to randomly select one reference user $q$ from the pool of $N_U^{(b)}$ users that have been randomly deployed in the beam coverage and to find the $K-1$ users that are closest to it in terms of the chosen similarity metric, \emph{i.e.}, by computing the distances $d_{qj} = \left.\parallel\mathbf{u}_{q}^{(b)}-\mathbf{u}_{j}^{(b)}\parallel\right.$, $\forall j$, and selecting the indexes corresponding to the $K$ lowest values (the lowest one is zero and corresponds to user $q$ by construction), \cite{Taricco,Prec14}. This algorithm is shown in Algorithm~\ref{alg:Bench_Upper}. After these users have been identified, in the following time frame, there is a new random selection of the reference user independently of whether it has already been served or not. Consequently, this algorithm does not satisfy the condition that the union set of all clusters provides the set of available users, $\bigcup_{i=1}^{N_K^{(b)}}\mathcal{C}_i^{(b)}\neq \mathcal{U}^{(b)}$, since $N_K^{(b)}=1$ and $K\neq N_U^{(b)}$. This is a strong assumption with a two-fold consequence: i) not all of the available users are served; and ii) the effect on the system performance of possible \emph{outliers}, \emph{i.e.}, users that are located far from all of the others in the considered space, is not considered. This algorithm  is thus always operating in a best-case scenario, since the reference user is always grouped with those closest to it. This is one of the main issues that a proper clustering algorithm shall solve, since the presence of outliers can significantly degrade the system performance, as will be clear after introducing the next algorithm that extends the one in \cite{Taricco,Prec14}. Based on these considerations, we refer to this algorithm as \emph{UpperBound} clustering.
\begin{algorithm}[t!]
 \caption{UpperBound clustering}
 \label{alg:Bench_Upper}
 \begin{algorithmic}[1]
 \renewcommand{\algorithmicrequire}{\textbf{Input:}}
 \renewcommand{\algorithmicensure}{\textbf{Output:}}
 \REQUIRE Feature vectors $\mathbf{u}_{j}^{(b)}$ and cluster size $K$, $\forall b=1,\ldots,N_B$, $j=1,\ldots,N_U^{(b)}$
 \ENSURE  One cluster per beam $\mathcal{C}^{(b)}$, $b=1,\ldots,N_B$
 \FOR {$b = 1$ to $N_B$}
     \STATE Randomly select one reference user $q$
     \STATE Compute:
     $d_{qj} = \left.\parallel\mathbf{u}_{q}^{(b)}-\mathbf{u}_{j}^{(b)}\parallel\right.$, $\forall j$
     \STATE Cluster together the $K$ users with lowest $d_{qi}$ values
      \RETURN $\mathcal{C}^{(b)}$
 \ENDFOR 
 \end{algorithmic} 
 \end{algorithm}
\subsubsection{Random clustering}
In \cite{UniLu_New}, the authors implement the same user selection algorithm as in the UpperBound, but at each time frame the reference user is randomly chosen among those that have not been served yet, \emph{i.e.}, for each beam all of the available users are clustered. As shown in Algorithm~\ref{alg:Fixed_A}, in step 7, the users that have not been yet allocated to a cluster, $\mathcal{Q}^{(b)}$, is updated and the algorithm continues to select a random user and to group it together with the closest $K-1$ users as long as $\mathcal{Q}^{(b)}$ is not empty (step 3). Thus, for each beam, a whole partition of the uniformly distributed users, $\mathcal{C}^{(b)} = \left\{\mathcal{C}^{(b)}_1,\ldots,\mathcal{C}^{(b)}_{N_K^{(b)}}\right\}$, is provided instead of one single cluster, which satisfies the condition $\bigcup_{i=1}^{N_K^{(b)}}\mathcal{C}_i^{(b)}= \mathcal{U}^{(b)}$. In the following, we refer to Algorithm~\ref{alg:Fixed_A} as to the Random algorithm, since the main characteristic to discern it from the next ones is actually given by the random choice of the reference user at each time frame.

\begin{figure}[!ht]
     \subfloat[Deployed users.\label{fig:Beam1_DeployedUsers}]{%
       \includegraphics[width=0.5\textwidth]{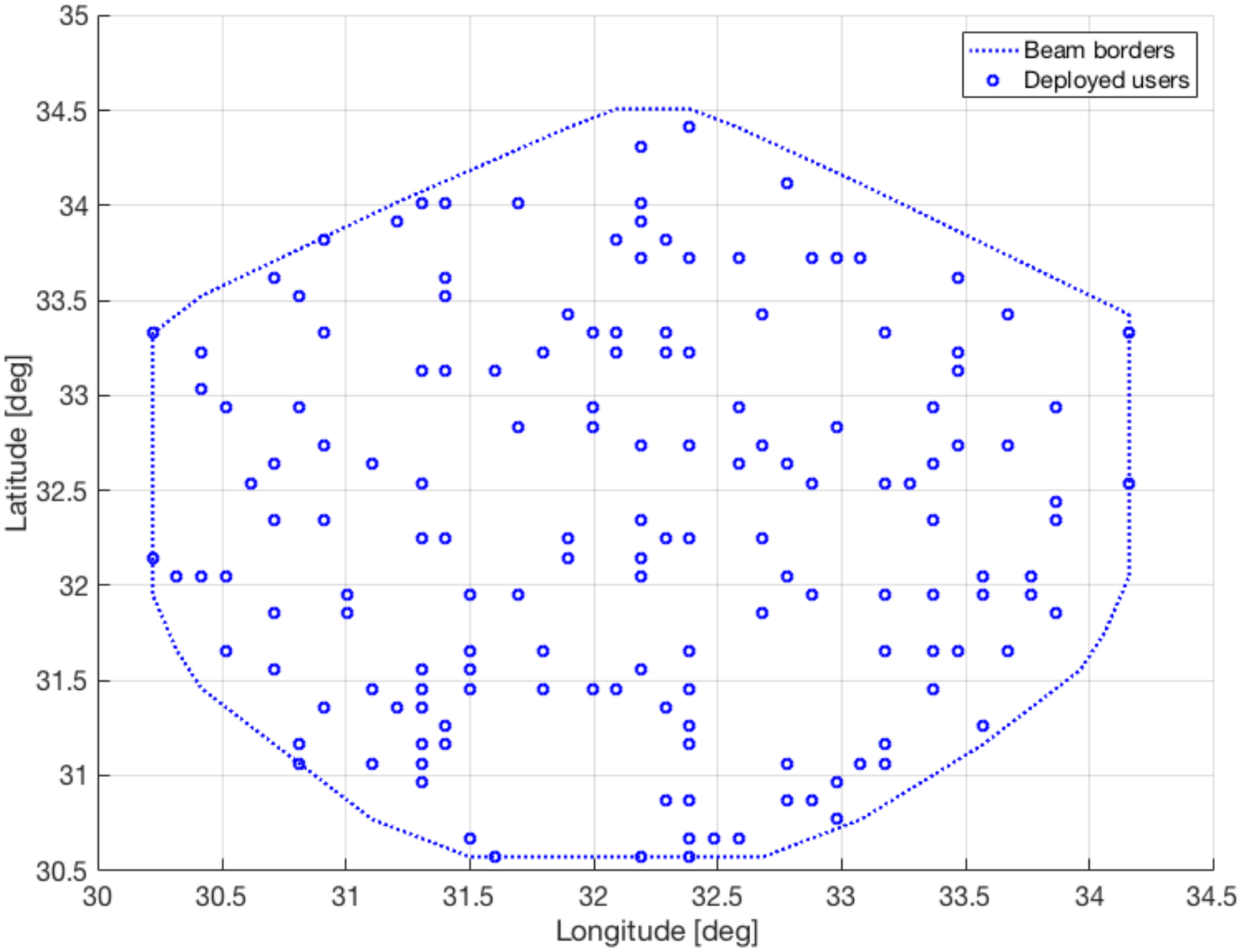}
     }
     \hfill
     \subfloat[Random.\label{fig:Beam1_Random}]{%
       \includegraphics[width=0.5\textwidth]{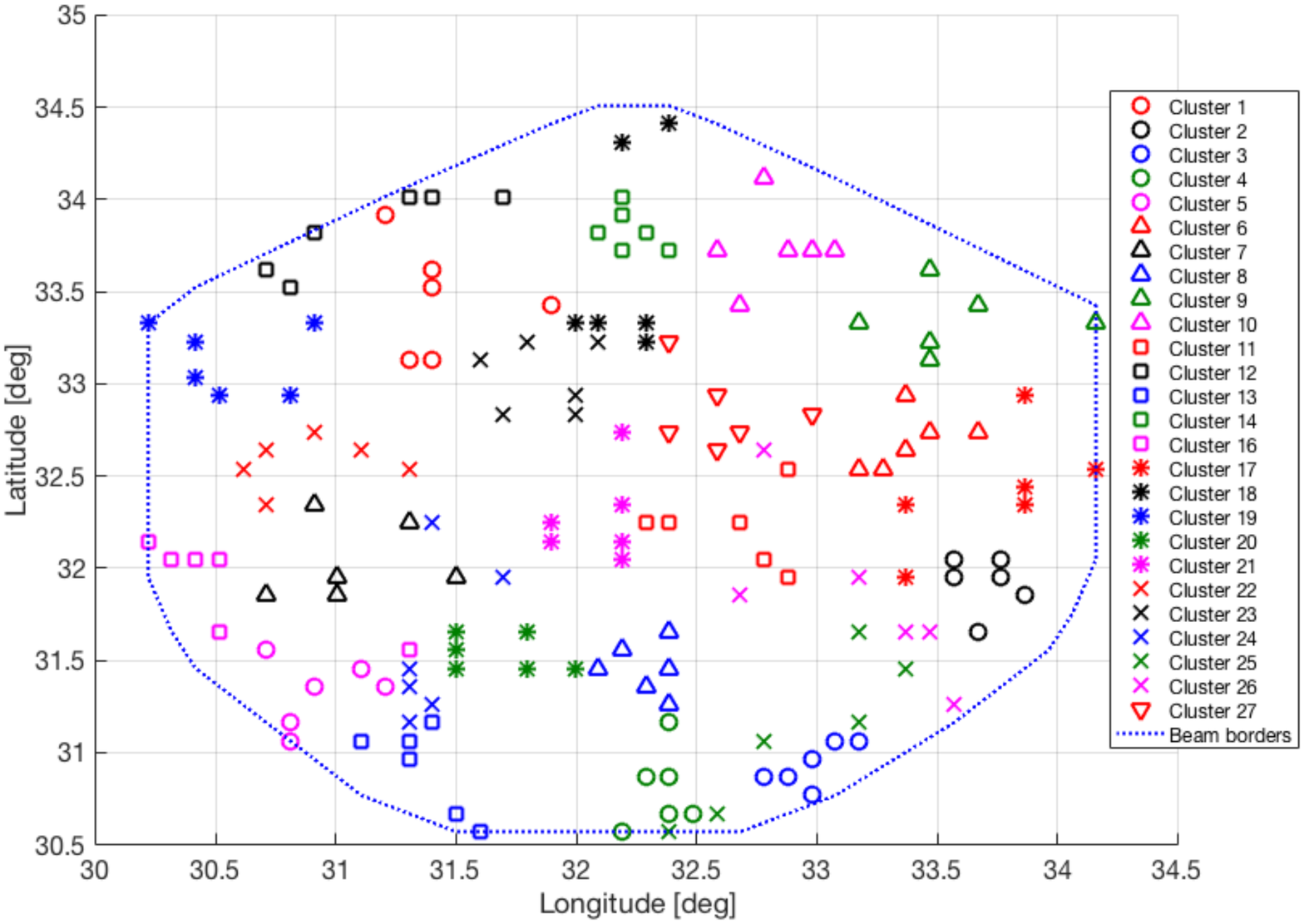}
     }
     \caption{Example of clustering with the Random algorithm. Setup: beam $1$, $\rho=1.25\cdot 10^{-3}$ users/km$^2$, $K=6$, Euclidean distance similarity.}
     \label{fig:dist_example}
\end{figure}

Figure~\ref{fig:Beam1_Random} shows a clustering example with the Random clustering algorithm. It can be noticed that there are some clusters that are formed by users located far apart. This happens when the reference user is an outlier, \emph{i.e.}, it is surrounded by already formed clusters and, thus, its only clustering possibility is given by distant users, as, for instance, clusters $18$ or $26$. The outliers problem can deeply degrade the overall system performance, since the farthest away the users in the similarity metric space, the less representative will be the average channel coefficient vector and, thus, the greater the SINR loss due to the cluster minimum SINR. In case the UpperBound algorithm is implemented, only cluster $1$ would be formed and served, thus not taking into account the outliers and providing a hardly achievable upper bound, as previously discussed.

\begin{algorithm}[t!]
 \caption{Random clustering}
 \label{alg:Fixed_A}
 \begin{algorithmic}[1]
 \renewcommand{\algorithmicrequire}{\textbf{Input:}}
 \renewcommand{\algorithmicensure}{\textbf{Output:}}
 \REQUIRE Feature vectors $\mathbf{u}_{j}^{(b)}$ and cluster size $K$, $\forall b=1,\ldots,N_B$, $j=1,\ldots,N_U^{(b)}$
 \ENSURE  $N_K^{(b)}$-partition $\mathcal{C}^{(b)}$, $\forall b$
 \FOR {$b = 1$ to $N_B$}
     \STATE Initialise the available users indexes $\mathcal{Q}^{(b)}=\left\{1,\ldots,N_U^{(b)}\right\}$ and the cluster index $c=1$
     \WHILE {$\mathcal{Q}^{(b)}\neq \emptyset$}
         \STATE Randomly select one reference user $q\in\mathcal{Q}^{(b)}$
         \STATE Compute:
     $d_{qj} = \left.\parallel\mathbf{u}_{q}^{(b)}-\mathbf{u}_{j}^{(b)}\parallel\right.$, $\forall j\in \mathcal{Q}^{(b)}$
         \STATE Cluster together the $K$ users with lowest $d_{qi}$ values in $\mathcal{C}_c^{(b)}$
         \STATE $\mathcal{Q}^{(b)} = \mathcal{Q}^{(b)}\setminus \mathcal{C}_c^{(b)}$, $c=c+1$
      \ENDWHILE
      \RETURN $\mathcal{C}^{(b)} = \left\{\mathcal{C}_1^{(b)},\ldots,\mathcal{C}_{N_K^{(b)}}^{(b)}\right\}$
 \ENDFOR 
 \end{algorithmic} 
 \end{algorithm}

\subsection{MaxDist}
The outliers problem in the Random algorithm is due to the random choice of a reference user around which the other cluster users are identified. This issue arises with increasing probability with the progress of the clustering procedure and it is more critical for low user densities, since the fewer and more spread the available users, the more likely they are to be far apart. Let us denote by $\mathbf{g}^{(b)}$ the barycentre of the users to be still allocated, $\mathcal{Q}^{(b)}$, in the considered space, \emph{i.e.}, $\mathbf{g}^{(b)} = \frac{1}{\left|\mathcal{Q}^{(b)}\right|}\sum_{j=1}^{\left|\mathcal{Q}^{(b)}\right|} \left.\parallel \mathbf{u}_j^{(b)} \right.\parallel$. From a geometrical point of view, the outliers problem arises when the distance of these users from the barycentre, $d_j =  \left.\parallel\mathbf{u}_{j}^{(b)}-\mathbf{g}^{(b)}\parallel\right.$, is large and, with a random selection as in Algorithm~\ref{alg:Fixed_A}, this scenario occurs with a non negligible probability. From an overall system performance point of view, it is desirable to first cluster the possible outliers so as to find the closest users in order to reduce the average intra-cluster distance. This can be achieved by selecting the reference user for each cluster based on the distance from the barycentre $\mathbf{g}^{(b)}$, as proposed in Algorithm~\ref{alg:MaxDist}. In the proposed algorithm, the reference user is no longer randomly chosen, but it is the user which is the farthest away from the available users barycentre $\mathbf{g}^{(b)}$, so that the first clusters to be generated are those for which the outliers problem is more critical. With this approach, the outliers in the system are clustered with the users that provide the best option for them in terms of similarity metric, while the last clusters are formed by those users that are the closest to the system barycentre, which guarantees by geometric construction that they show good similarity conditions. Clearly, the barycentre $\mathbf{g}^{(b)}$ shall be updated at each iteration based on the set of available users $\mathcal{Q}^{(b)}$.

An example of the proposed algorithm, denoted as \emph{MaxDist} algorithm to highlight the different reference user selection, is provided in Figure~\ref{fig:Beam1_MaxDist}. With respect to the Random clustering in Figure~\ref{fig:Beam1_Random}, it can be noticed that the first clusters to be formed are indeed composed by the users that were outliers Algorithm~\ref{alg:Fixed_A}. In particular, while with the Random algorithm there are users of clusters $7$ and $8$ in the left-most part of the beam that are very distant from the other cluster members, with the proposed MaxDist approach, these users are now both allocated to cluster $1$, thus limiting the outliers problem. Finally, we can also notice that even the last clusters in Figure~\ref{fig:Beam1_MaxDist} do not suffer from the presence of outliers, since the last users to be grouped together are, by geometrical construction, those in the best conditions in terms of similarity to the others.
\begin{algorithm}[t!]
 \caption{MaxDist clustering}
 \label{alg:MaxDist}
 \begin{algorithmic}[1]
 \renewcommand{\algorithmicrequire}{\textbf{Input:}}
 \renewcommand{\algorithmicensure}{\textbf{Output:}}
 \REQUIRE Feature vectors $\mathbf{u}_{j}^{(b)}$ and cluster size $K$, $\forall b=1,\ldots,N_B$, $j=1,\ldots,N_U^{(b)}$
 \ENSURE  $N_K^{(b)}$-partition $\mathcal{C}^{(b)}$, $\forall b$
 \FOR {$b = 1$ to $N_B$}
     \STATE Initialise the available users indexes $\mathcal{Q}^{(b)}=\left\{1,\ldots,N_U^{(b)}\right\}$ and the cluster index $c=1$
     \WHILE {$\mathcal{Q}^{(b)}\neq \emptyset$}
         \STATE Compute the barycentre of the available users $\mathbf{g}^{(b)} = \frac{1}{\left|\mathcal{Q}^{(b)}\right|}\sum_{j=1}^{\left|\mathcal{Q}^{(b)}\right|} \left.\parallel \mathbf{u}_j^{(b)} \right.\parallel$
         \STATE Select $q$ such that $\left.\parallel\mathbf{u}_{q}^{(b)}-\mathbf{g}^{(b)}\parallel\right.\geq \left.\parallel\mathbf{u}_{j}^{(b)}-\mathbf{g}^{(b)}\parallel\right.$, $\forall j\in\mathcal{Q}^{(b)}$
         \STATE Compute:
     $d_{qj} = \left.\parallel\mathbf{u}_{q}^{(b)}-\mathbf{u}_{j}^{(b)}\parallel\right.$, $\forall j\in \mathcal{Q}^{(b)}$
         \STATE Cluster together the $K$ users with lowest $d_{qi}$ values in $\mathcal{C}_c^{(b)}$
         \STATE $\mathcal{Q}^{(b)} = \mathcal{Q}^{(b)}\setminus \mathcal{C}_c^{(b)}$, $c=c+1$
      \ENDWHILE
      \RETURN $\mathcal{C}^{(b)} = \left\{\mathcal{C}_1^{(b)},\ldots,\mathcal{C}_{N_K^{(b)}}^{(b)}\right\}$
 \ENDFOR 
 \end{algorithmic} 
 \end{algorithm}

\subsection{kmeans++}
Clustering algorithms have been explored in the past with respect to both internal homogeneity (\emph{i.e.}, similarity among cluster members) and external separation (\emph{i.e.}, dissimilarity with respect to other clusters), \cite{Book1,Book2,Book3,kmeans1,kmeansPP}, and they can be classified into: i) \emph{partitional clustering}, which aim at defining a one-level partitioning of the initial objects into non-overlapping and non-empty clusters by means of maximisation (minimisation) of a given cost function; and ii) \emph{hierarchical clustering}, aiming at identifying a tree-like partition with a single, all-inclusive cluster at the top and the single initial objects at the bottom as singleton clusters. The Random and MaxDist algorithms discussed in the previous paragraphs can be broadly considered fixed-size partitional clustering algorithms, although the definition of the required partition is not iterated so as to minimise/maximise a cost function. In the following, we focus on partitional clustering algorithms that minimise a cost function $\mathcal{J}$ based on the distance in the considered similarity multi-dimensional space. In particular, we consider a Sum-of-Squared-Error (SSE) cost function, which for the $b$-th beam is defined as:
\begin{equation}
\label{eq:SSEcost}
\begin{split}
    \mathcal{J}^{(b)}\left(\bm{\Lambda}^{(b)},\mathbf{M}^{(b)}\right) &= \sum_{i=1}^{N_K^{(b)}}\sum_{j=1}^{N_U^{(b)}} \lambda_{ij}^{(b)} {\left.\parallel \mathbf{u}_j^{(b)}-\mathbf{m}_i^{(b)} \parallel\right.}^2 \\
    &= \sum_{i=1}^{N_K^{(b)}}\sum_{j=1}^{N_U^{(b)}} \mathcal{J}_{ij}^{(b)}
\end{split}
\end{equation}
where: i) $\mathbf{M}^{(b)}={\left({\left(\mathbf{m}_1^{(b)}\right)}^T,\ldots,{\left(\mathbf{m}_{N_K^{(b)}}^{(b)}\right)}^T\right)}^T$ is the centroid (prototype) matrix in which the $i$-th row is the barycentre of the users belonging to the $i$-th cluster, \emph{i.e.}, $\mathbf{m}_i^{(b)} = \frac{1}{\left|\mathcal{C}_i^{(b)}\right|}\sum_{j=1}^{\left|\mathcal{C}_i^{(b)}\right|} \lambda_{ij}^{(b)}\mathbf{u}_j^{(b)}$; ii) $\bm{\Lambda}^{(b)}$ is a $N_K^{(b)}\times N_U^{(b)}$ partition matrix in which $\lambda_{ij}^{(b)}=1$ is $1$ iff $\mathbf{u}_j^{(b)}\in\mathcal{C}_i^{(b)}$ and $0$ otherwise; and iii) $\mathcal{J}_{ij}^{(b)}=\lambda_{ij}^{(b)} {\left.\parallel \mathbf{u}_j^{(b)}-\mathbf{m}_i^{(b)} \parallel\right.}^2$ is the cost function evaluated for the $i$-th object and $j$-th centroid. It is worthwhile noting that the partition that minimises the above cost function is optimal and leads to the \emph{minimum variance partition} with respect to the chosen similarity metric.

\begin{algorithm}[t!]
\caption{k-means++ clustering}
 \label{alg:kmeans}
 \begin{algorithmic}[1]
 \renewcommand{\algorithmicrequire}{\textbf{Input:}}
 \renewcommand{\algorithmicensure}{\textbf{Output:}}
 \REQUIRE Feature vector $\mathbf{u}_j^{(b)}$ and number of clusters per beam $N_K^{(b)}$, $\forall j,b$
 \ENSURE  $N_K^{(b)}$-partition $\mathcal{C}^{(b)}=\left\{\mathcal{C}_1^{(b)},\ldots,\mathcal{C}_{N_K^{(b)}}^{(b)}\right\}$, $\forall b$
 \STATE k-means++ centroids initialisation $\mathbf{M}^{(b)}={\left({\left(\mathbf{m}_1^{(b)}\right)}^T,\ldots,{\left(\mathbf{m}_{N_K^{(b)}}^{(b)}\right)}^T\right)}^T$
  \FOR {$b = 1$ to $N_B$}
  \WHILE{$\mathbf{M}^{(b)}$ is varying}
      \FOR {$j=1,\ldots,N_U^{(b)}$}
          \IF {$\left(\left.\parallel\mathbf{u}_j^{(b)}-\mathbf{m}_{\ell}^{(b)}\parallel\right.<\left.\parallel\mathbf{u}_j^{(b)}-\mathbf{m}_{i}^{(b)}\parallel\right.\right)$, $i=1,\ldots,N_K^{(b)}$}
          \STATE Assign the object to the nearest cluster: $\mathbf{u}_j^{(b)}\in\mathcal{C}_{\ell}^{(b)}\Rightarrow \lambda_{\ell j}^{(b)} = 1$
          \ENDIF
      \ENDFOR
      \STATE Update the centroid vectors $\mathbf{m}_i^{(b)} = \frac{1}{\left|\mathcal{C}_i^{(b)}\right|}\sum_{j=1}^{N_U^{(b)}} \lambda_{ij}^{(b)}\mathbf{u}_j^{(b)}$.
      \ENDWHILE
       \RETURN $\mathcal{C}^{(b)}=\left\{\mathcal{C}_1^{(b)},\ldots,\mathcal{C}_{N_K^{(b)}}^{(b)}\right\}$
  \ENDFOR
 \end{algorithmic} 
 \end{algorithm}
 
One of the earliest algorithms developed within this framework is the \emph{k-means} algorithm, \cite{kmeans1}, which is based on an iterative approach that seeks a $N_K^{(b)}$-partition of the input data sets, starting from a randomly chosen set of initial centroids, by minimising the defined SSE cost function. The main drawback of the  traditional k-means algorithm is that its performance strongly depends on the choice of the initial centroids, which has two critical consequences: i) the worst-case running time of the algorithm is super-polynomial in the input size; and ii) the clustering result can be arbitrarily bad with respect to the cost function compared to the optimal clustering. Several solutions have been proposed in the literature to improve the initial centroids selection, \cite{kmeans2,kmeans3}. Notably, the best performing solution is the \emph{k-means++} algorithm in which the initial centroid selection is based on a probabilistic approach, \cite{kmeansPP}. In particular, the first centroid is randomly chosen from the input set $\mathcal{U}^{(b)}$ with uniform distribution, while the remaining $N_{K}^{(b)}-1$ initial centroids are selected from the remaining ones with a probability proportional to the squared distance from their closest existing cluster center. In the following, we focus on the k-means++ algorithm since numerical simulations show that this implementation can perform twice as fast as the traditional k-means and also provide significantly better clustering solutions.

\begin{figure}[!ht]
     \subfloat[MaxDist.\label{fig:Beam1_MaxDist}]{%
       \includegraphics[width=0.5\textwidth]{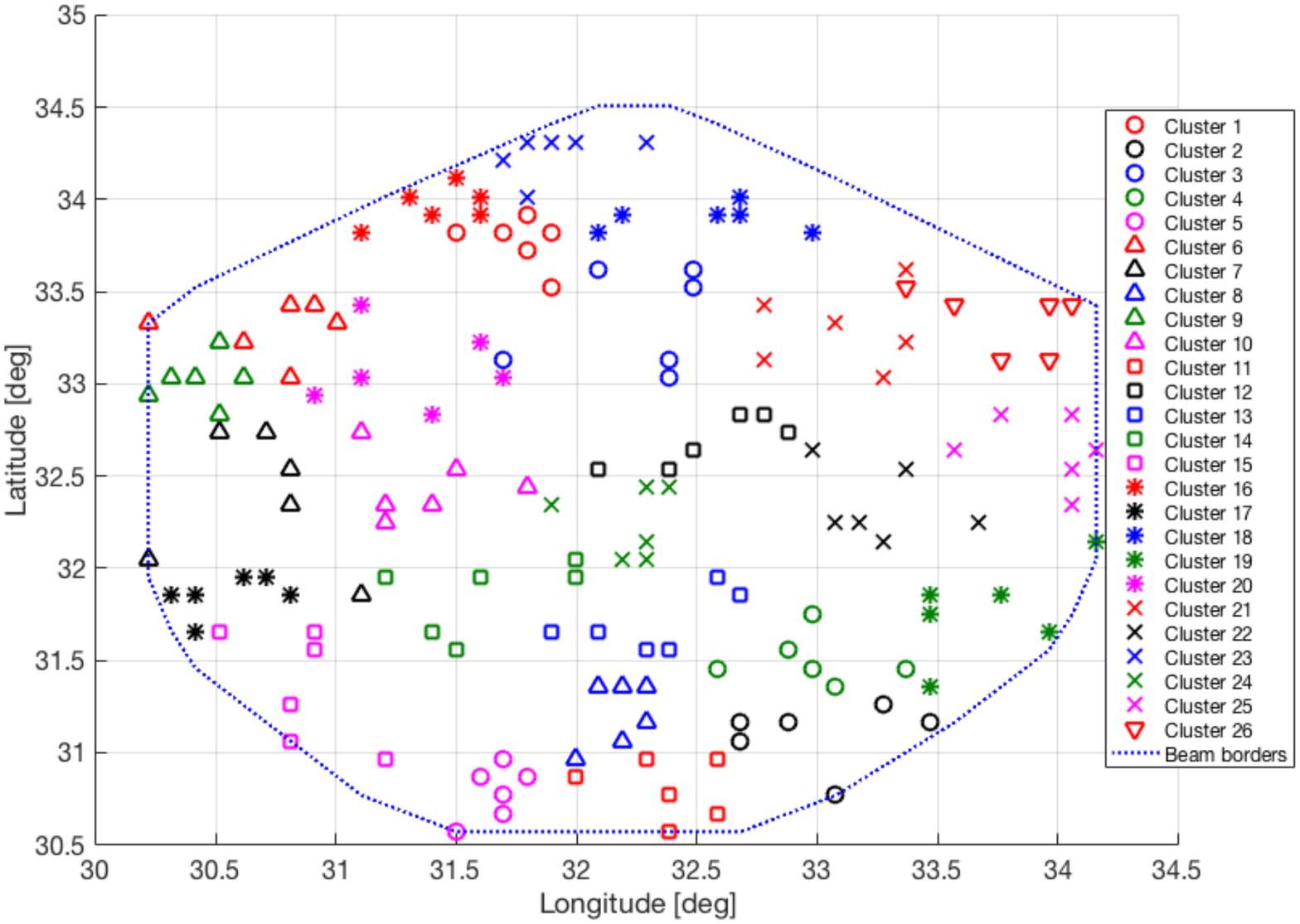}
     }
     \hfill
     \subfloat[kmeans++.\label{fig:Beam1_kmeans}]{%
       \includegraphics[width=0.5\textwidth]{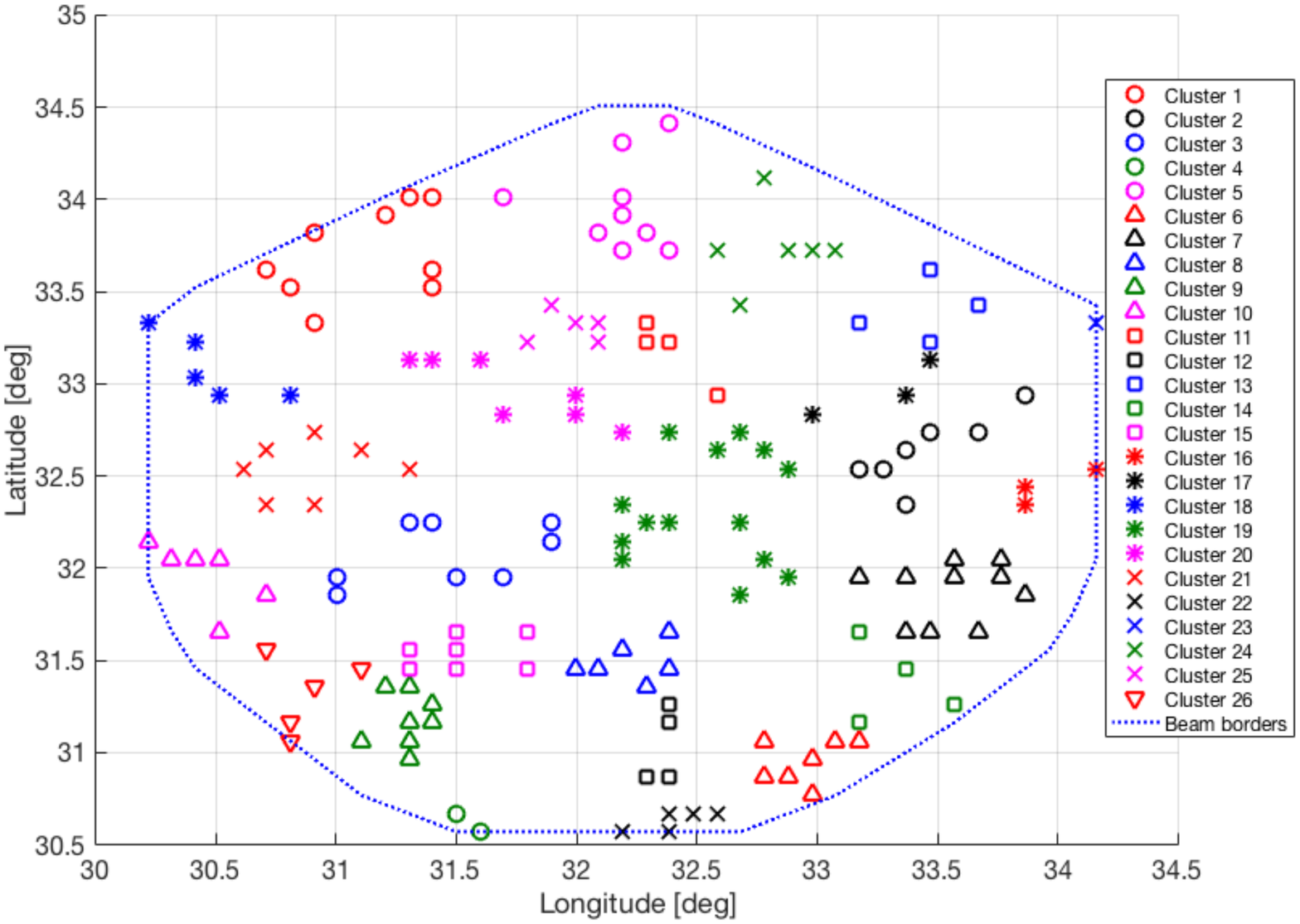}
     }
     \caption{Example of clustering with the MaxDist and kmeans++ algorithms. Setup: beam $1$, $\rho=1.25\cdot 10^{-3}$ users/km$^2$, $K=6$, Euclidean distance similarity.}
     \label{fig:dist_example}
\end{figure}

An extremely important difference with respect to fixed size clustering algorithms, as the Random and MaxDist ones considered in this work, is that the k-means++ algorithm does not provide a fixed-size partition. The iterative algorithm keeps updating the clusters composition so as to find the partition that minimises the cost function in (\ref{eq:SSEcost}), as shown in steps 3 and 9 in Algorithm~\ref{alg:kmeans}. Thus, the k-means++ algorithm does not take a fixed cluster size $K$ as input, while it requires the number of clusters, $N_K^{(b)}$. It is straightforward to show that this aspect implies that the k-means++ algorithm will provide a $N_K^{(b)}$-partition with an \emph{average cluster size} $K$, \emph{i.e.}, the average number of users per cluster within a given beam, equal to $K$. If we denote by $K_c^{(b)}$ the $c$-th cluster size in the $b$-th beam, $c=1,\ldots,N_K^{(b)}$, it can be shown that $\left(1/N_K^{(b)}\right)\sum_{c=1}^{N_K^{(b)}} K_b^{(c)} = K$. For each beam to be served, we can now implement the k-means++ algorithm to identify the users grouping required by the multicast precoder as shown in Algorithm~\ref{alg:kmeans}. It is worth noting that step $5$ involves the nearest neighbour rule, in the space defined by the objects features, and it is, thus, a Voronoi tessellation, \cite{Voronoi1}. An example of the k-means++ clustering is given in Figure~\ref{fig:Beam1_kmeans}. It can be noticed that the average cluster size is equal to the fixed cluster size in Figure~\ref{fig:Beam1_Random} and Figure~\ref{fig:Beam1_MaxDist}, but the dimension of each cluster is now varying so as to obtain a minimum variance partition.
 
\begin{figure}[!t]
    \centering
    \includegraphics[width=0.6\columnwidth]{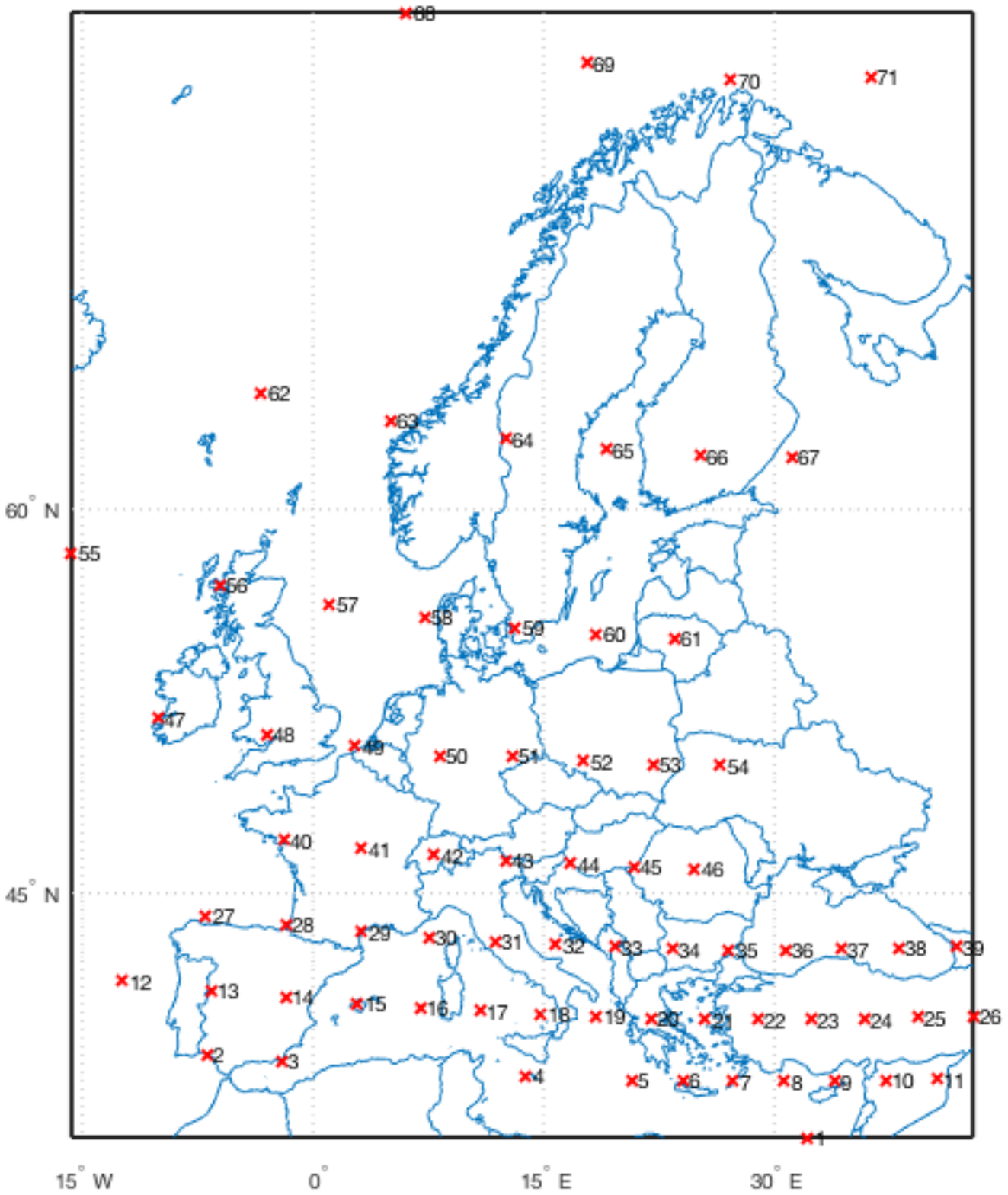}
    \vspace{-25mm}
    \caption{Multi-beam satellite system covering Europe through $71$ beams. The red crosses identify the beam centers.}
    \label{fig:multi-beamPlot}
\end{figure}

\section{Numerical Results}
\label{sec:Results}
In this section, we assess the performance of the clustering algorithms described in Section~\ref{sec:Clustering} and compare them in terms of average spectral efficiency, fairness, and complexity. The considered multi-beam satellite system is shown in Fig.~\ref{fig:multi-beamPlot} and it covers the whole of Europe through $N_B=71$ beams operating with a full frequency reuse scheme. The main simulation parameters are listed in Table~\ref{tab:SimulationParam}. With respect to the considered user densities, the largest value is $10^{-2}$ users/km$^2$, which is a low density. While this choice is motivated by the memory and time computational complexity, it will be shown that the trend of the overall system for increasing densities is already evident with the considered values. During each Monte Carlo iteration, $N_U^{(b)}=\left[\rho A_b\right]$ users are randomly deployed in fixed locations in each beam with a uniform distribution, as previously shown in Fig.~\ref{fig:Beam1_DeployedUsers}, and the considered clustering algorithms are implemented to obtain a fixed- (UpperBound, Random, and MaxDist) or variable- (k-means++) size partition. We further assume that the traffic request from the users is uniform, \emph{i.e.}, all users are requesting the same amount of traffic and no priorities are present or requested. Moreover, as previously highlighted, different beams will have a different number of clusters. Consequently, when a certain beam already served all of its users but other beams still are not complete, it randomly picks an already served cluster to serve it again. This assumption does not affect the validity of the proposed analysis due to the uniform traffic assumption. In case non-uniform traffic distributions are considered, this shall be properly reflected in the scheduler. 
 
For the generic $c$-th cluster in the $b$-th beam and $n$-th Monte Carlo iteration, we obtain a rate $\eta_{c,n}^{(b)}\left(K,\rho\right)$\footnote{The same formulation applies to the k-means++ algorithm, the only difference being in the interpretation of $K$.}, which is a function of the minimum SINR among the cluster members, \emph{i.e.}:
\begin{equation}
\label{eq:NumericalRate}
    \eta_{c,n}^{(b)}(K,\rho) = f\left(\widetilde{\gamma}_c^{(b)}\right),\ \widetilde{\gamma}_c^{(b)}=\min_{i\in\mathcal{C}_c^{(b)}}\left\{\gamma_{c,i}^{(b)}\right\}
\end{equation}
The function $f(\cdot)$ models the considered Modulation and Coding (ModCod) scheme, which in the following is assumed to be the one provided by the DVB-S2X standard with $64800$ bits FEC codewords, \cite{DVBS2X}. The above minimum SINR depends on the similarity metric used to describe the users feature space, the (average) cluster size, and the user density, since these are the design parameters that affect the performance of the proposed algorithm. In the following, we provide a thorough insight on the impact that each of these parameters has on the overall system performance, \emph{i.e.}, the average rate. This value is obtained by averaging over all iterations, simulated clusters, and number of beams:
\begin{equation}
\label{eq:AverageRate}
    \overline{\eta}(K,\rho) = \mathbb{E}_{n,c,b}\left\{\eta_{c,n}^{(b)}(K,\rho)\right\}
\end{equation}
The following numerical results have been obtained with PAC and SPC precoding and with both Euclidean and channel coefficient similarity metrics. It shall be noted that, although the maximum considered user density is limited because of computational complexity, the considered user density range already provides a clear indication of the relation between this parameter and the considered key performance indicators.

\begin{table}[t]
\renewcommand{\arraystretch}{1.3}
\caption{Numerical simulation parameters}
\label{tab:SimulationParam}
\centering
\begin{tabular}{|c|c|}
\hline
\bfseries Parameter & \bfseries Value\\
\hline\hline
Carrier frequency & $19.5$ GHz\\
\hline
Receiving antenna diameter & $0.6$ m\\
\hline
Receiving antenna efficiency & $0.6$\\
\hline
Antenna losses & $2.55$ dB\\
\hline
GEO satellite longitude & $30^{\circ}$\\
\hline
Satellite transmitted power & $P_{sat}=90$W\\
\hline
$\rho$ & $1.25\cdot 10^{-3}, 2.5\cdot 10^{-3}, 10^{-2}$ users/km$^2$\\
\hline
$K$ & $1$ (unicast), $2,4,\ldots,12$\\
\hline
Target Bit Error Probability & $10^{-5}$\\
\hline
\end{tabular}
\end{table}

\subsection{Average Rate}
\begin{figure}[!t]
     \subfloat[Euclidean distance.\label{fig:CAPACITY_DIST_PAC_L}]{%
       \includegraphics[width=0.5\textwidth]{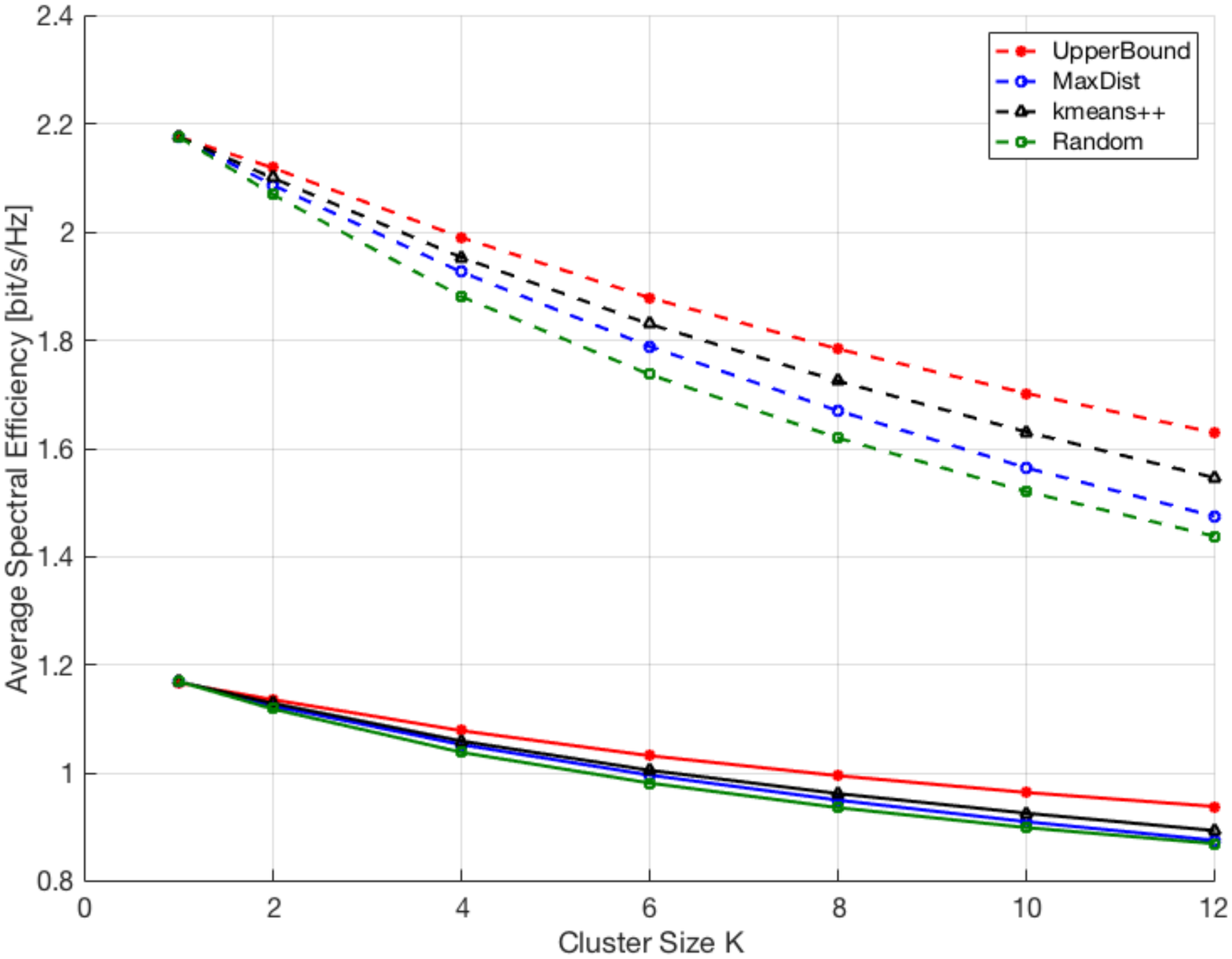}
     }
     \hfill
     \subfloat[Channel coefficients.\label{fig:CAPACITY_CHANN_PAC_L}]{%
       \includegraphics[width=0.5\textwidth]{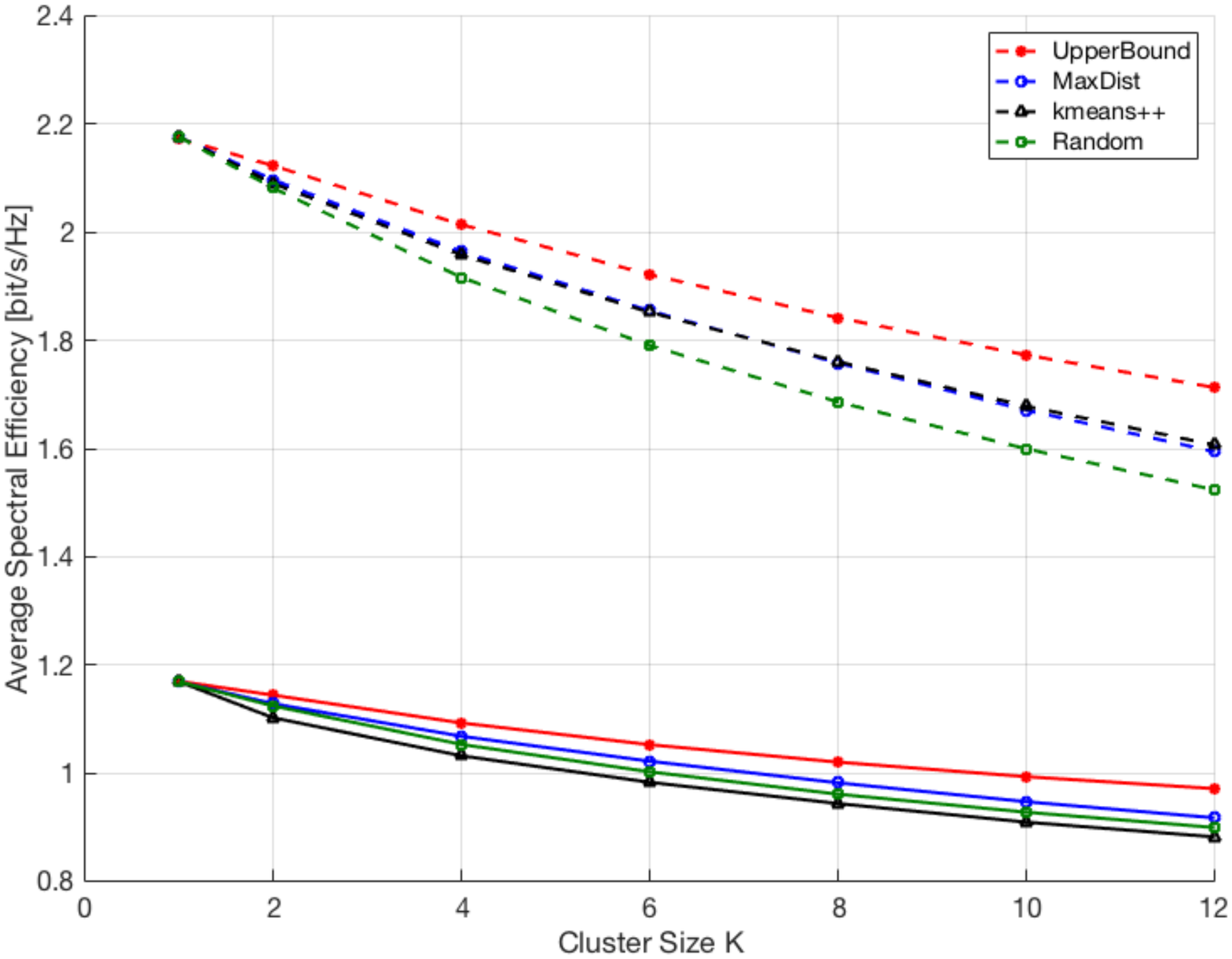}
     }
     \caption{Average rate per beam with (dashed line) and without MMSE PAC precoding. Setup: $\rho = 1.25\cdot 10^{-3}$ users/km$^2$.}
     \label{fig:Rate_PAC_L}
\end{figure}

\begin{figure}[!t]
     \subfloat[Euclidean distance.\label{fig:CAPACITY_DIST_PAC_M}]{%
       \includegraphics[width=0.5\textwidth]{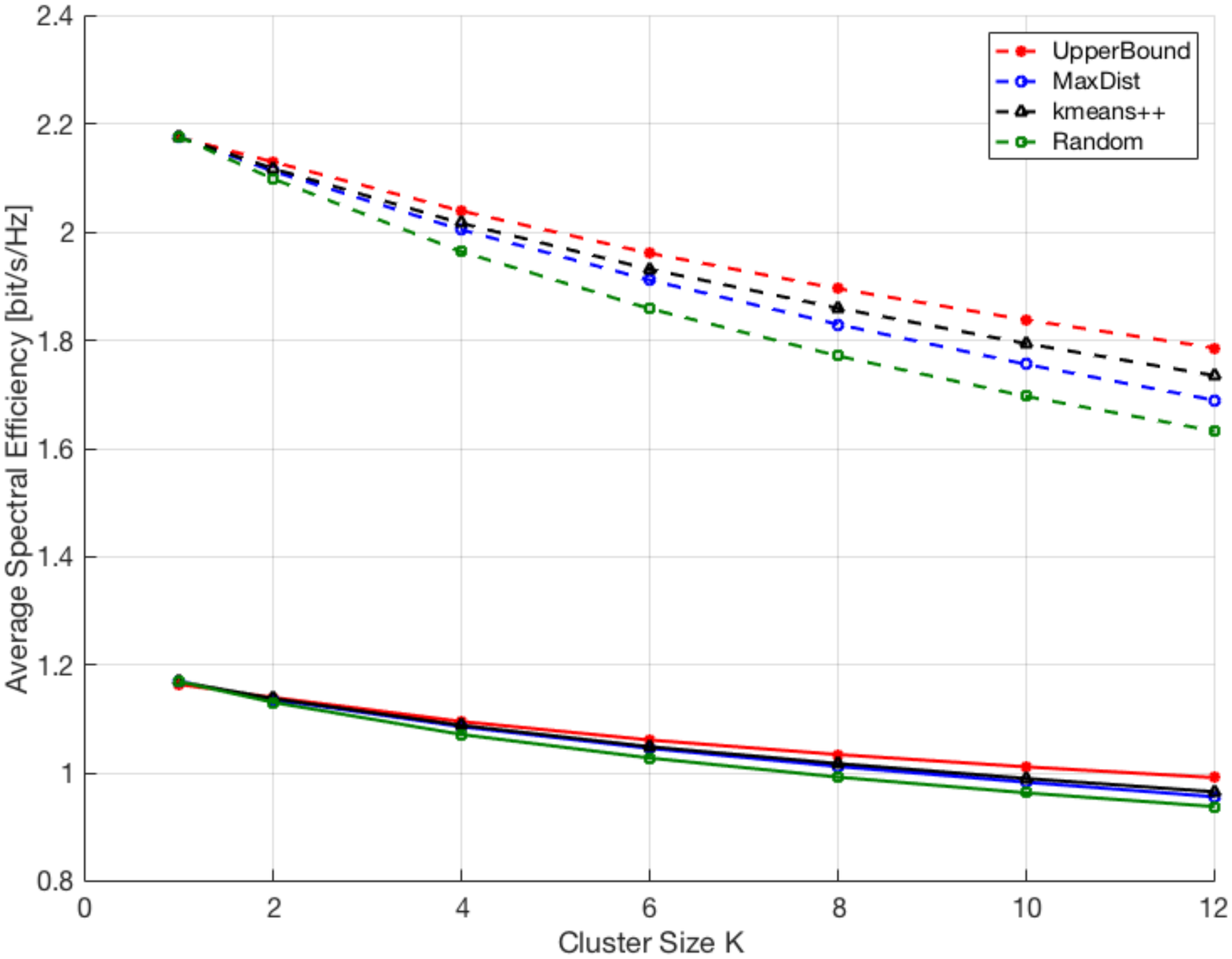}
     }
     \hfill
     \subfloat[Channel coefficients.\label{fig:CAPACITY_CHANN_PAC_M}]{%
       \includegraphics[width=0.5\textwidth]{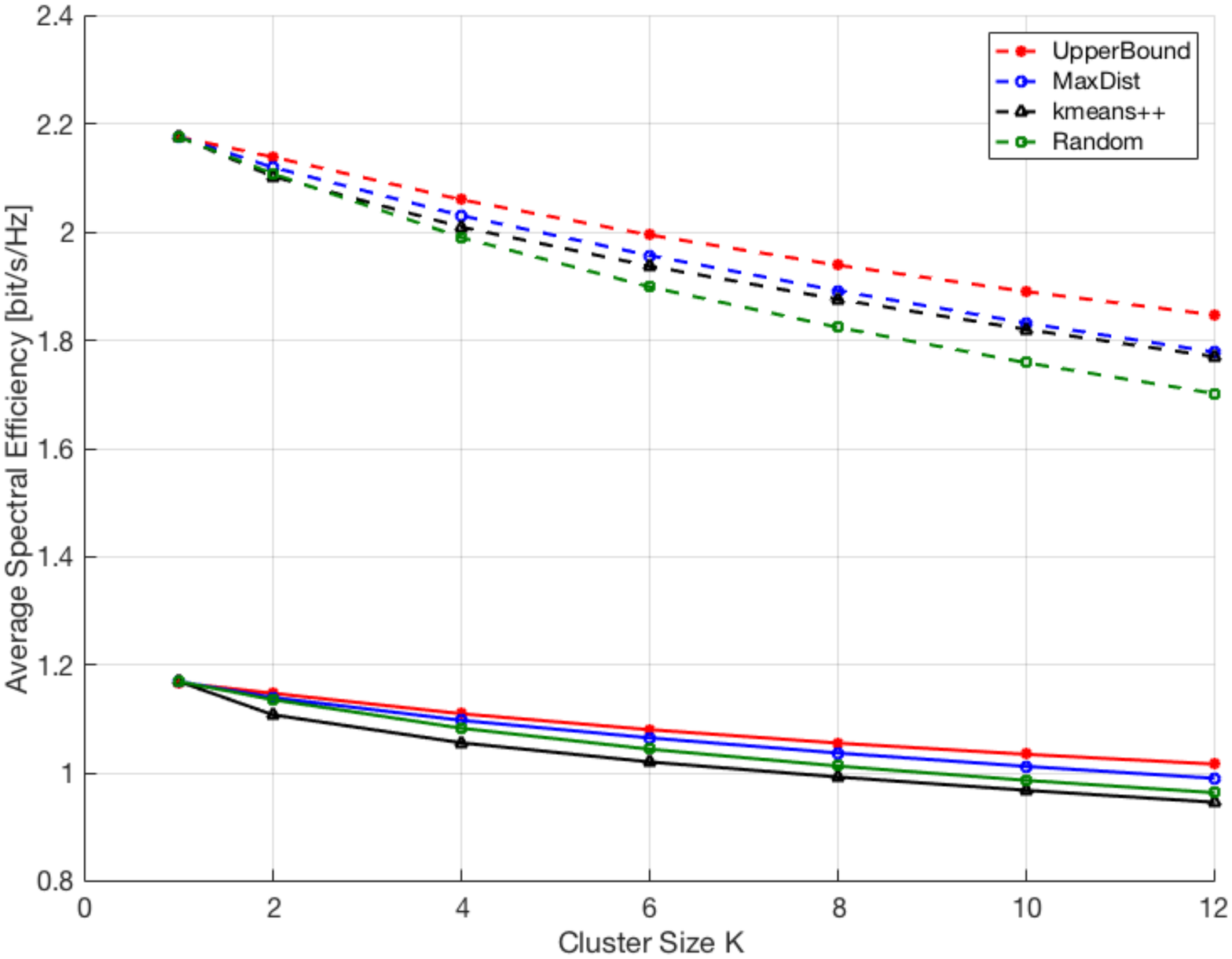}
     }
     \caption{Average rate per beam with (dashed line) and without MMSE PAC precoding. Setup: $\rho = 2.5\cdot 10^{-3}$ users/km$^2$.}
     \label{fig:Rate_PAC_M}
\end{figure}

\begin{figure}[!t]
     \subfloat[Euclidean distance.\label{fig:CAPACITY_DIST_PAC_H}]{%
       \includegraphics[width=0.5\textwidth]{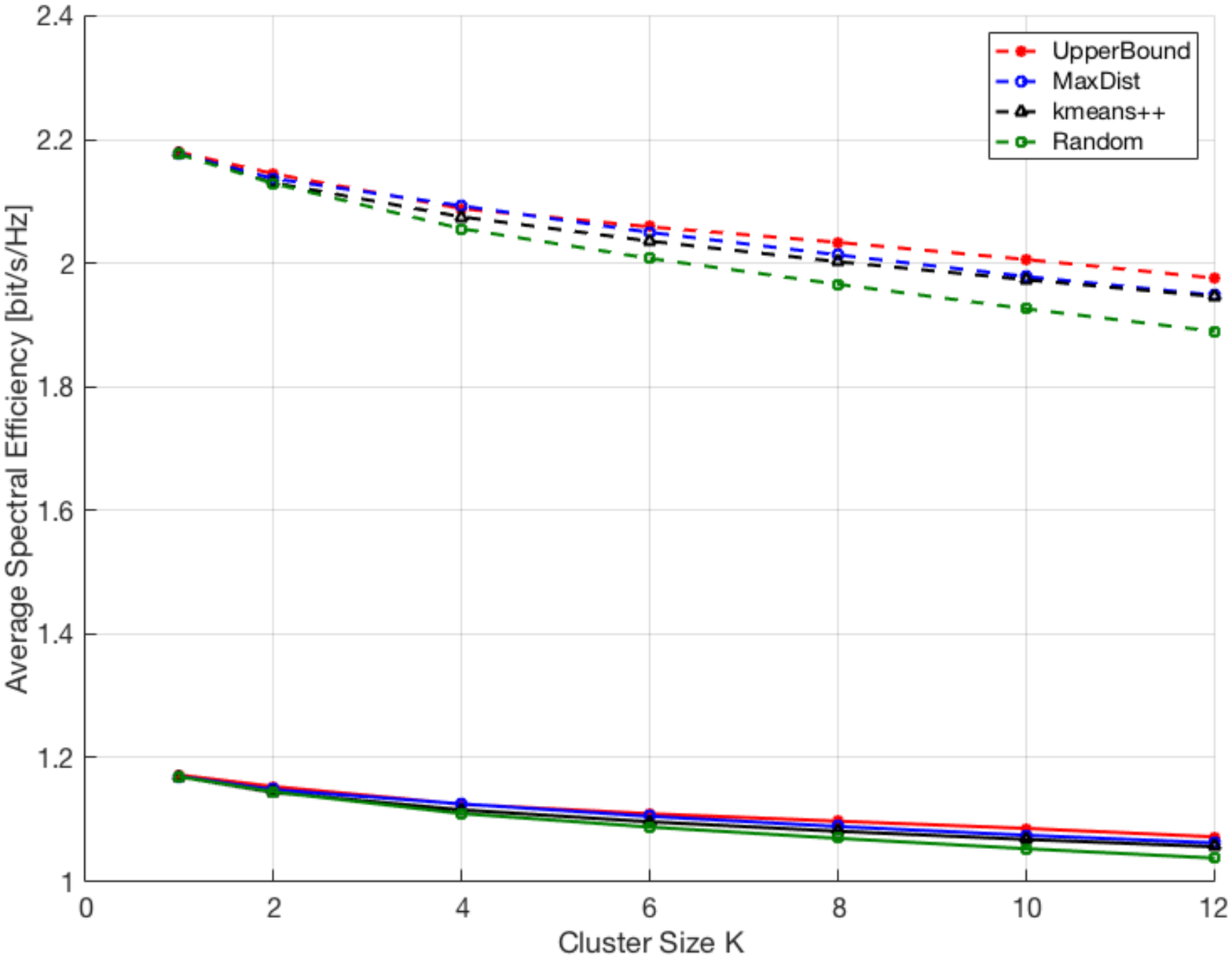}
     }
     \hfill
     \subfloat[Channel coefficients.\label{fig:CAPACITY_CHANN_PAC_H}]{%
       \includegraphics[width=0.5\textwidth]{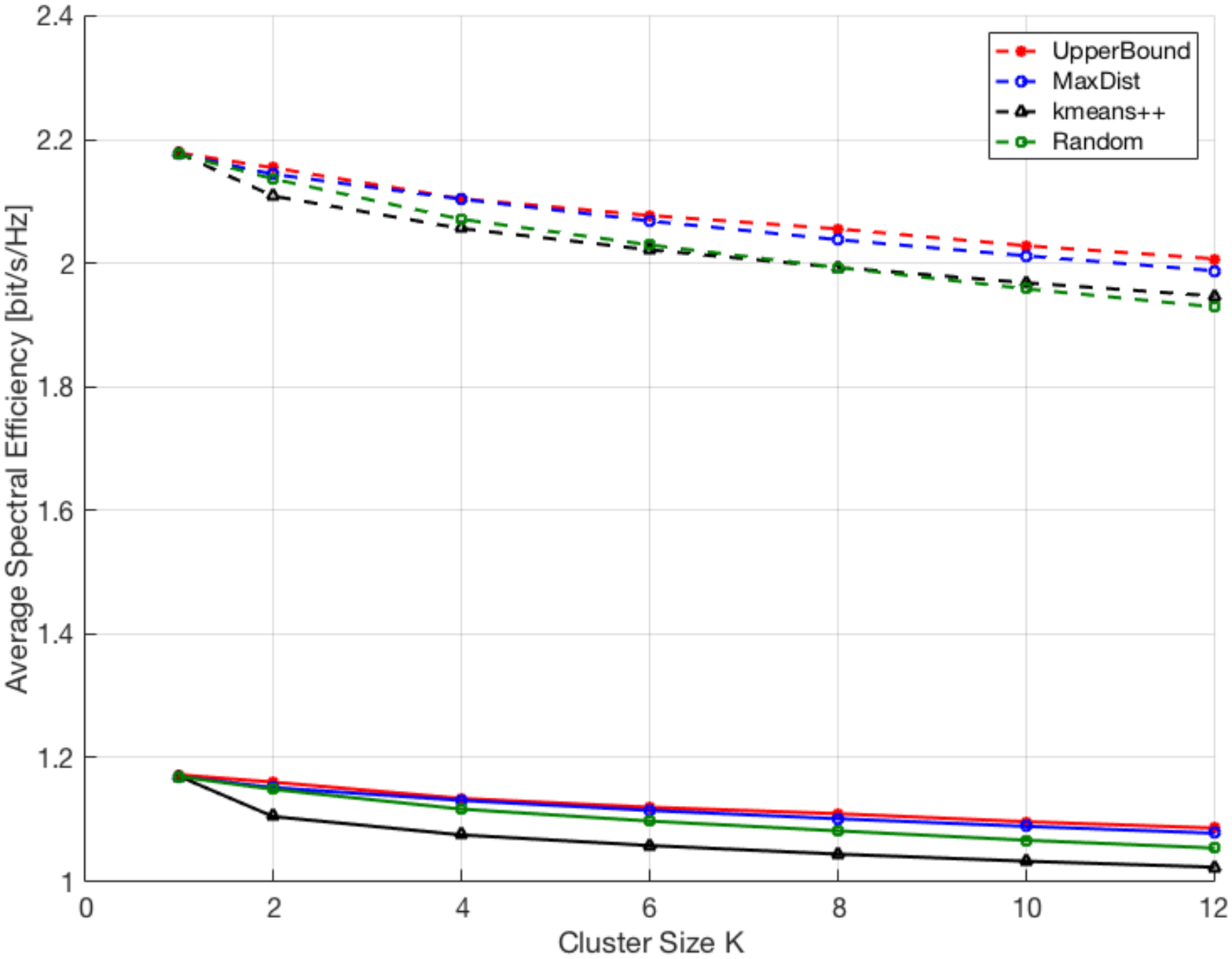}
     }
     \caption{Average rate per beam with (dashed line) and without MMSE PAC precoding. Setup: $\rho = 10^{-2}$ users/km$^2$.}
     \label{fig:Rate_PAC_H}
\end{figure}

\begin{figure}[!t]
     \subfloat[Euclidean distance.\label{fig:CAPACITY_DIST_SPC_L}]{%
       \includegraphics[width=0.5\textwidth]{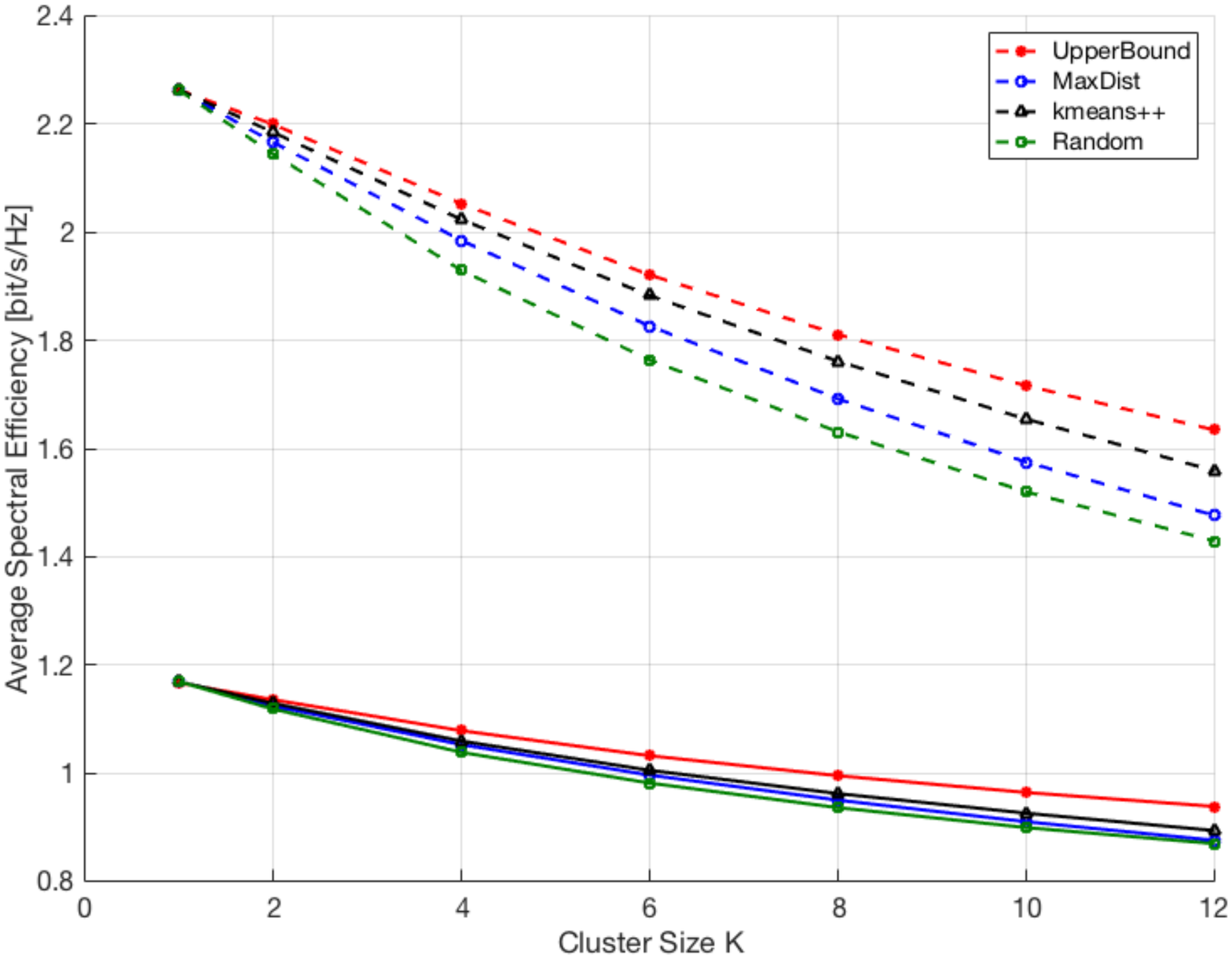}
     }
     \hfill
     \subfloat[Channel coefficients.\label{fig:CAPACITY_CHANN_SPC_L}]{%
       \includegraphics[width=0.5\textwidth]{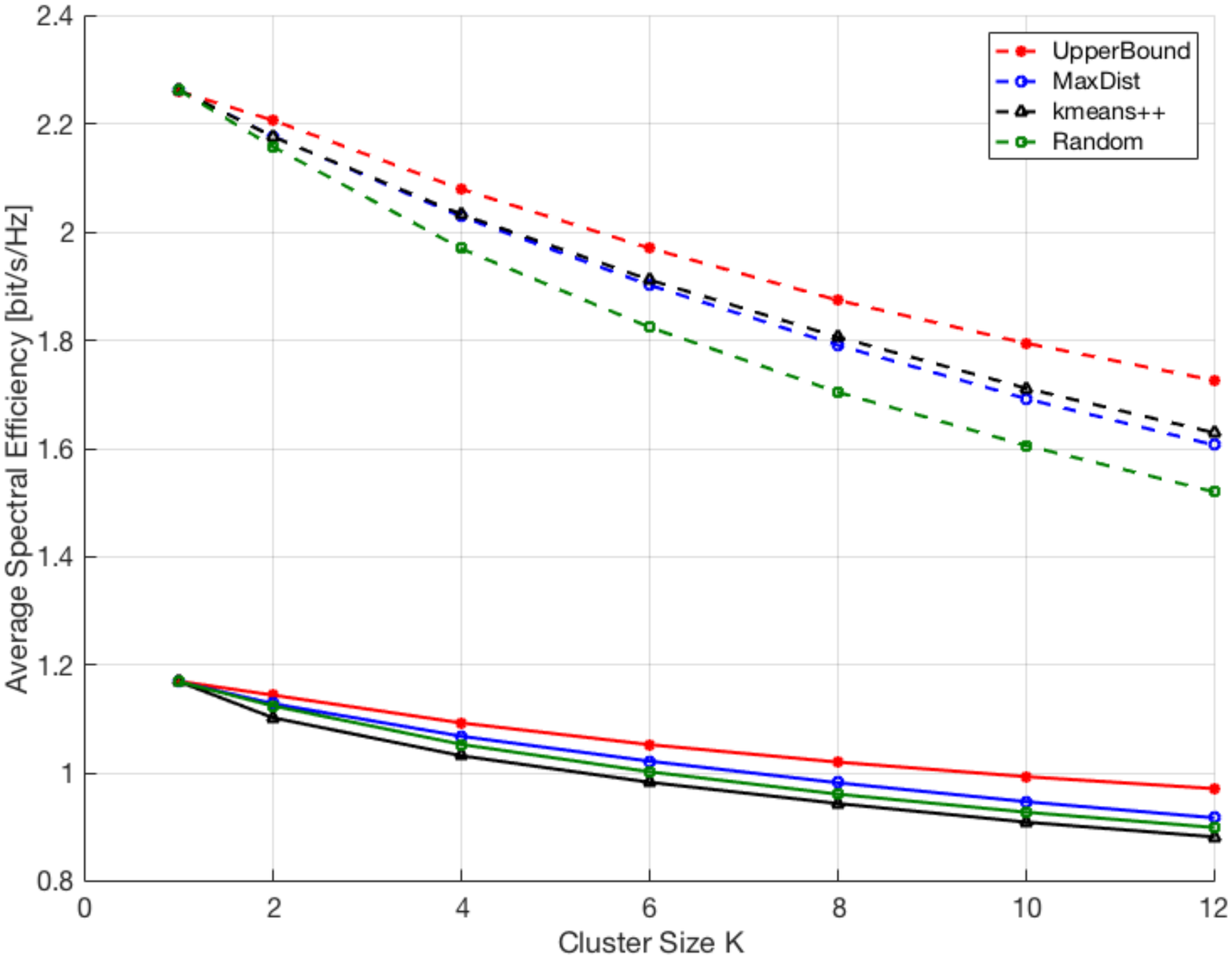}
     }
     \caption{Average rate per beam with (dashed line) and without MMSE SPC precoding. Setup: $\rho = 1.25\cdot 10^{-3}$ users/km$^2$.}
     \label{fig:Rate_SPC_L}
\end{figure}

\begin{figure}[!t]
     \subfloat[Euclidean distance.\label{fig:CAPACITY_DIST_SPC_M}]{%
       \includegraphics[width=0.5\textwidth]{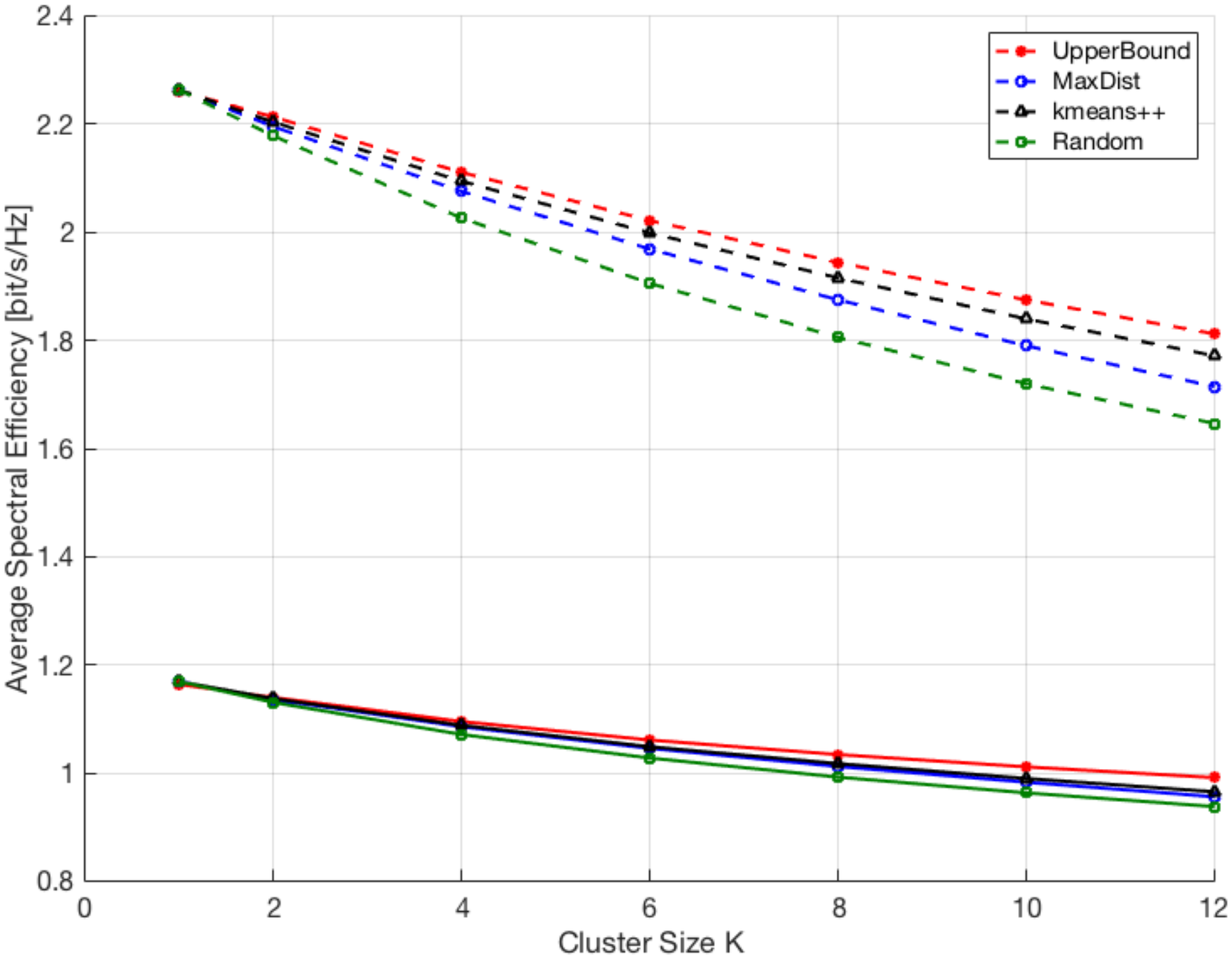}
     }
     \hfill
     \subfloat[Channel coefficients.\label{fig:CAPACITY_CHANN_SPC_M}]{%
       \includegraphics[width=0.5\textwidth]{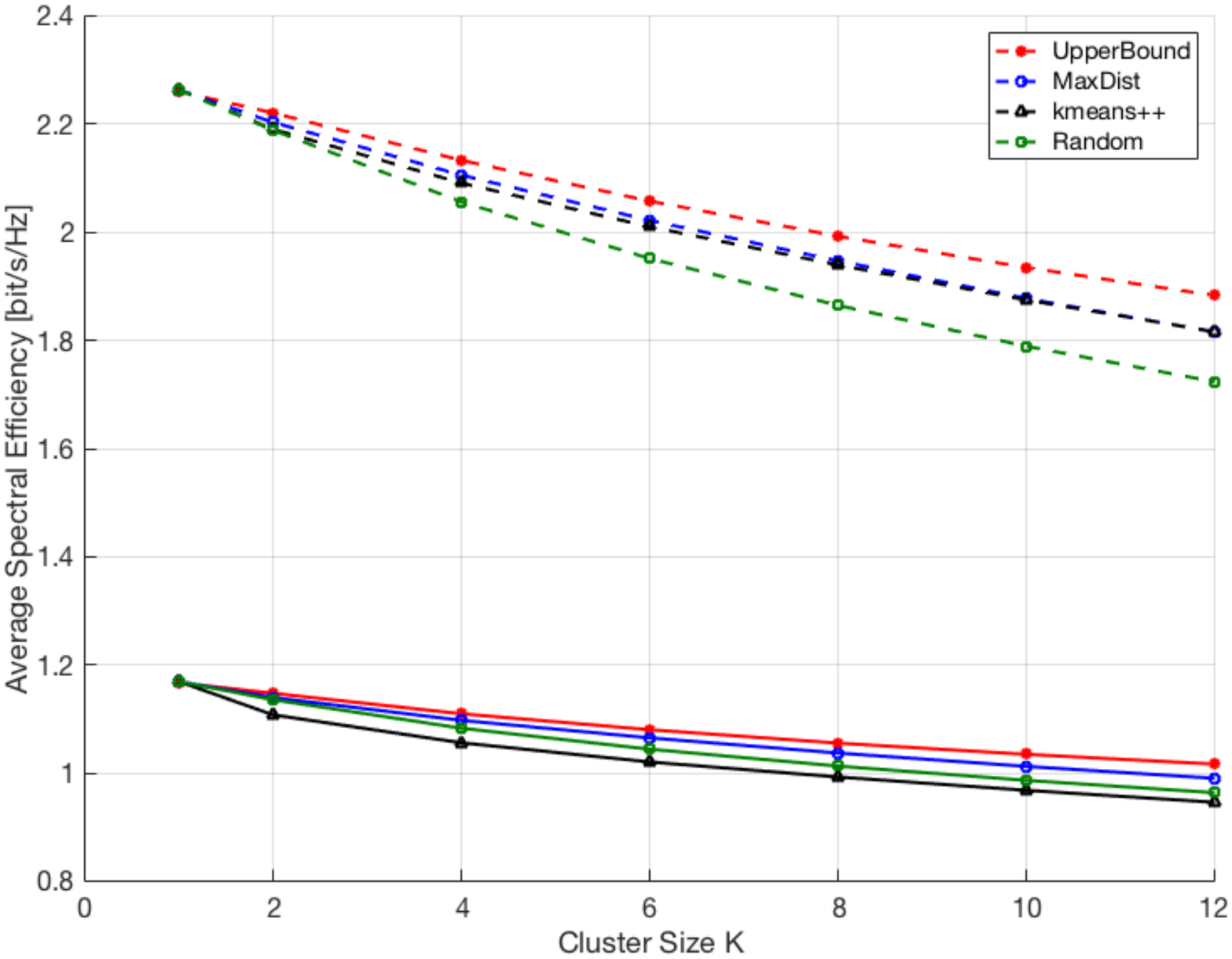}
     }
     \caption{Average rate per beam with (dashed line) and without MMSE SPC precoding. Setup: $\rho = 2.5\cdot 10^{-3}$ users/km$^2$.}
     \label{fig:Rate_SPC_M}
\end{figure}

\begin{figure}[!t]
     \subfloat[Euclidean distance.\label{fig:CAPACITY_DIST_SPC_H}]{%
       \includegraphics[width=0.5\textwidth]{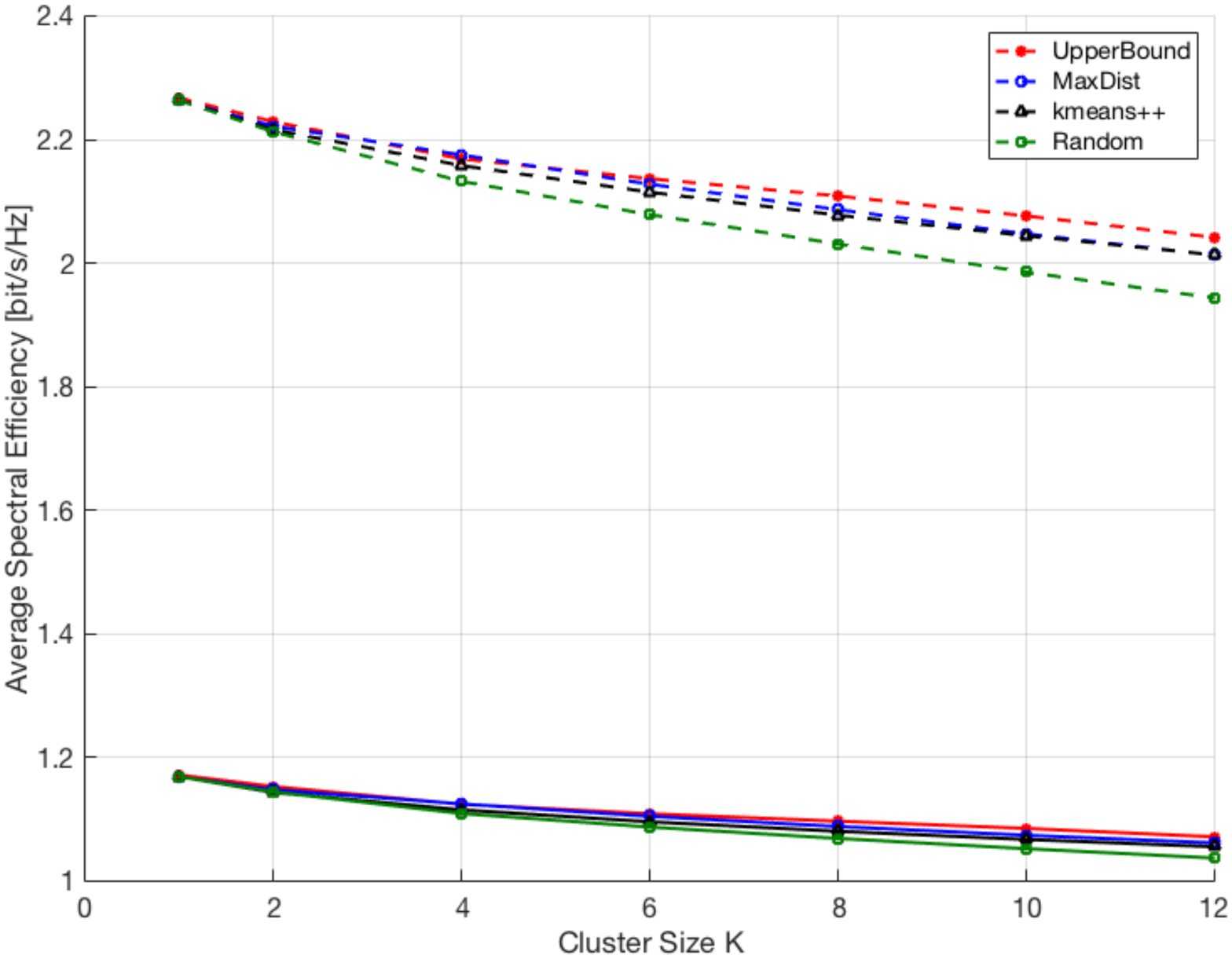}
     }
     \hfill
     \subfloat[Channel coefficients.\label{fig:CAPACITY_CHANN_SPC_H}]{%
       \includegraphics[width=0.5\textwidth]{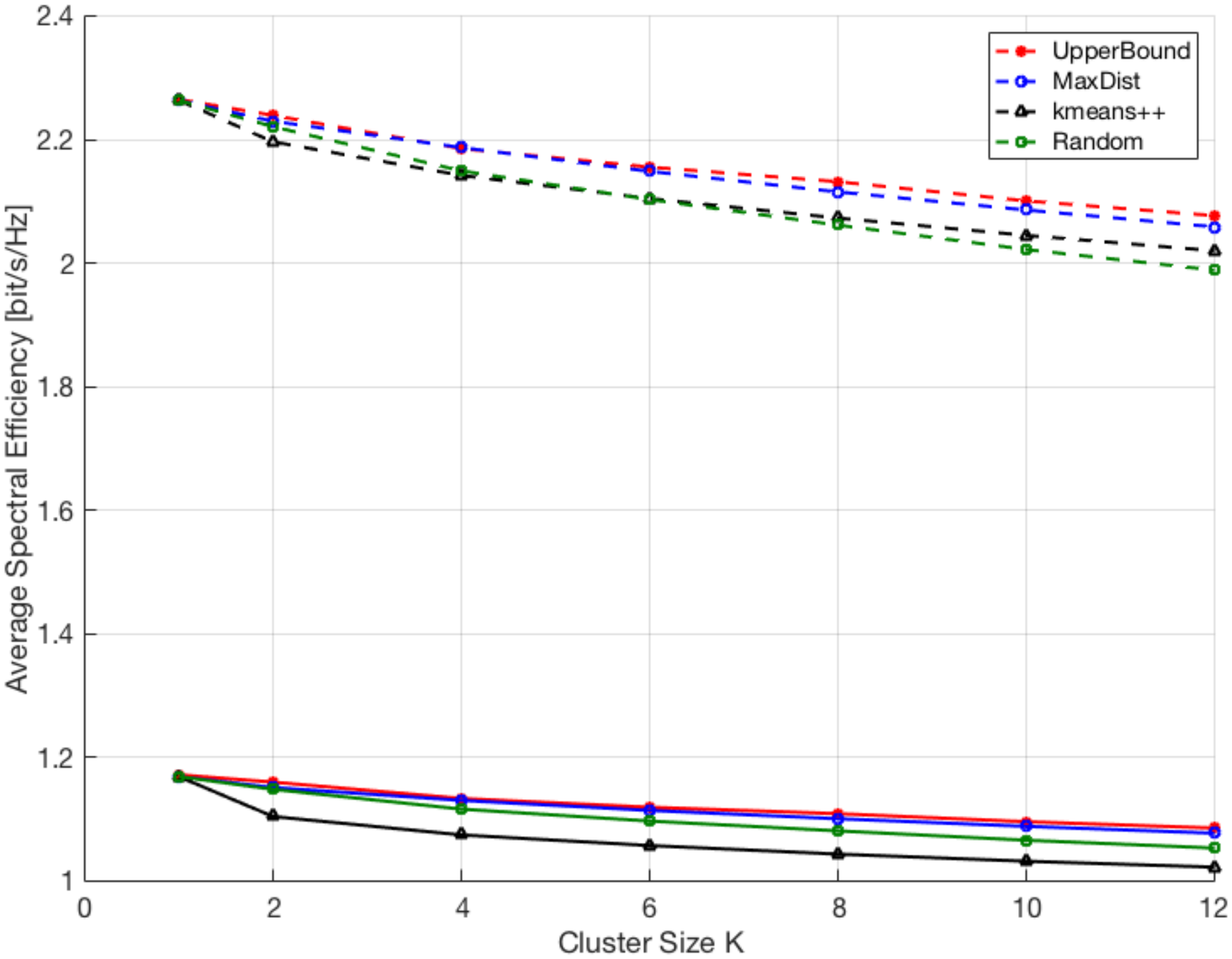}
     }
     \caption{Average rate per beam with (dashed line) and without MMSE SPC precoding. Setup: $\rho = 10^{-2}$ users/km$^2$.}
     \label{fig:Rate_SPC_H}
\end{figure}

Figures~\ref{fig:Rate_PAC_L}-\ref{fig:Rate_SPC_H} provide the average rate defined in eq.~(\ref{eq:AverageRate}) as a function of the cluster size $K$ (which is to be intended as the average cluster size for the k-means++ algorithm), both with and without PAC or SPC MMSE precoding. First of all, it is worth highlighting that the average rate tends to decrease also when not implementing precoding for increasing values of the cluster size. This behaviour can be easily explained by observing that, even without precoding, users are still clustered together based on the considered clustering algorithm and, thus, also in this case there is a loss related to the user with lowest SINR. The implementation of precoding techniques provides a significant improvement in terms of average rate.

For all of the analysed clustering algorithms, both with PAC and SPC precoding, the average rate $\overline{\eta}(K,\rho)$ decreases for increasing values of the cluster size and this effect is more relevant with lower user densities. This is due to the fact that, by increasing $K$, more users with different channel conditions are grouped into the same FEC codeword. Since the precoding matrix in eq.~(\ref{eq:MMSE_PREC}) is built by averaging the channel vectors of the users within a cluster, when more users are clustered together in low density spaces the resulting MMSE precoding vector is poorly adapted to the actual channel experienced by the users and the performance is more degraded. When considering larger values of the user density (see Figures \ref{fig:Rate_PAC_H} and \ref{fig:Rate_SPC_H} for $\rho=0.01$ users/km$^2$), the loss for increasing cluster size $K$ is lower: in this case, in fact, the clusterisation algorithms can actually find users that are closer to each other with both the Euclidean distance and the channel coefficients similarity metrics.

% RATE L ----------------------------------------------------------------
% RATE M ----------------------------------------------------------------
% RATE H ----------------------------------------------------------------

% RATE L ----------------------------------------------------------------
% RATE M ----------------------------------------------------------------
% RATE H ----------------------------------------------------------------

\begin{figure}[!t]
    \centering
    \includegraphics[width=0.5\columnwidth]{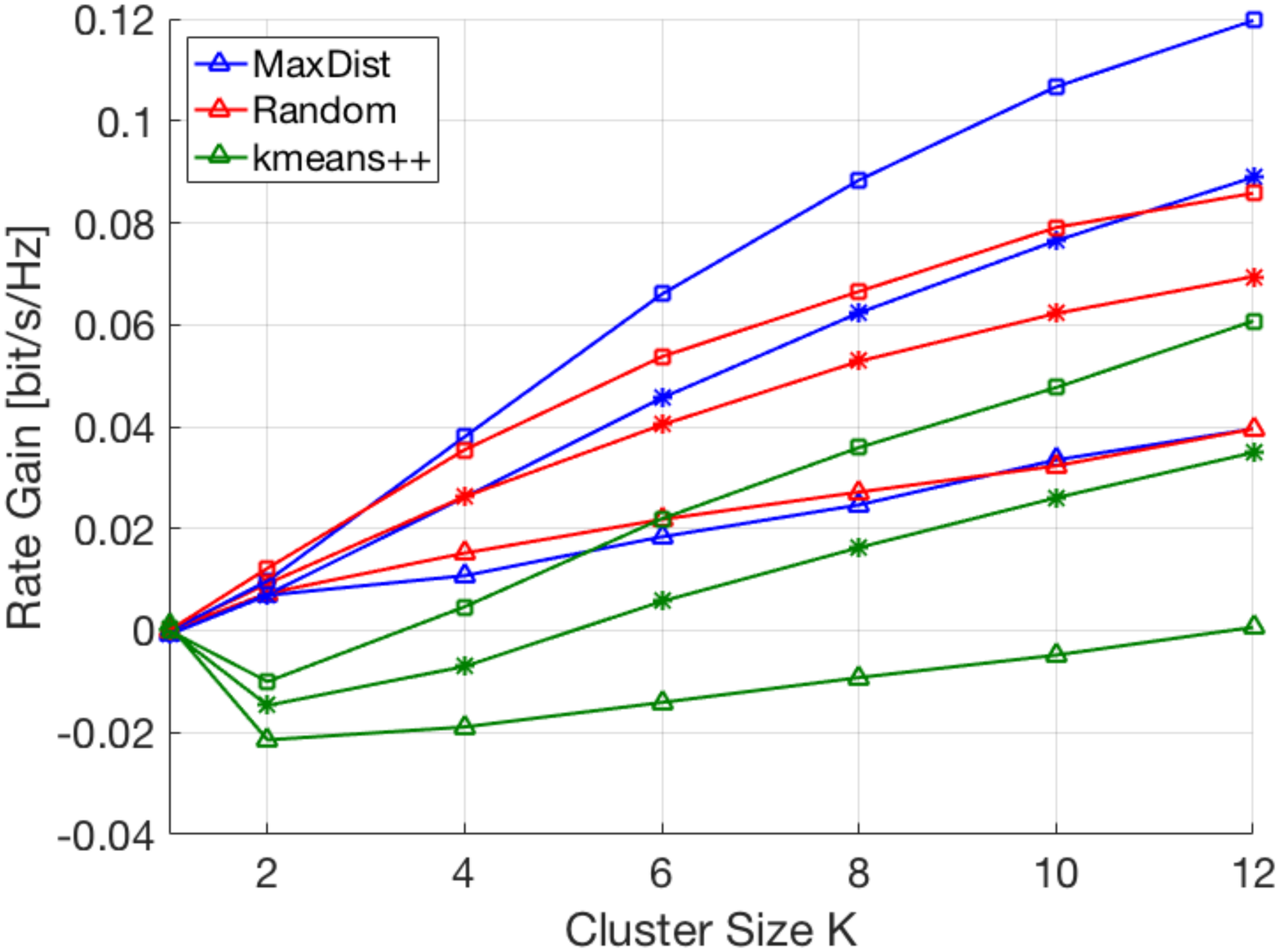}
    \caption{Difference in average spectral efficiency versus cluster size between the channel coefficient and the Euclidean distance similarity metrics. Setup: $\rho=0.01$ users/km$^2$ (triangle), $\rho=2.5\cdot 10^{-3}$ users/km$^2$ (star), $\rho=1.25\cdot 10^{-3}$ users/km$^2$ (square).}
    \label{fig:RateGain}
\end{figure}

\begin{figure}[!t]
     \subfloat[Euclidean distance.]{%
       \includegraphics[width=0.5\textwidth]{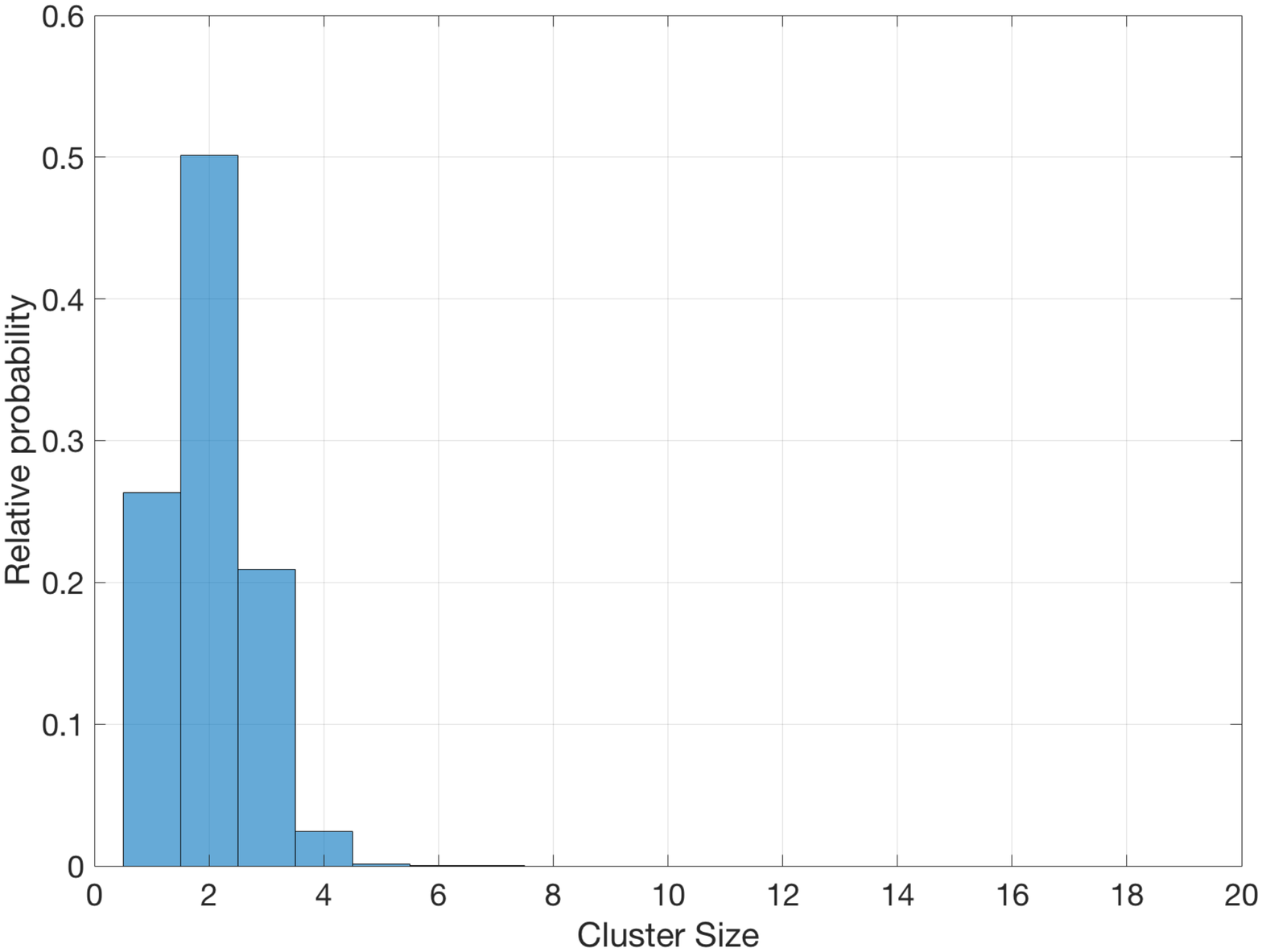}
     }
     \hfill
     \subfloat[Channel coefficients.]{%
       \includegraphics[width=0.5\textwidth]{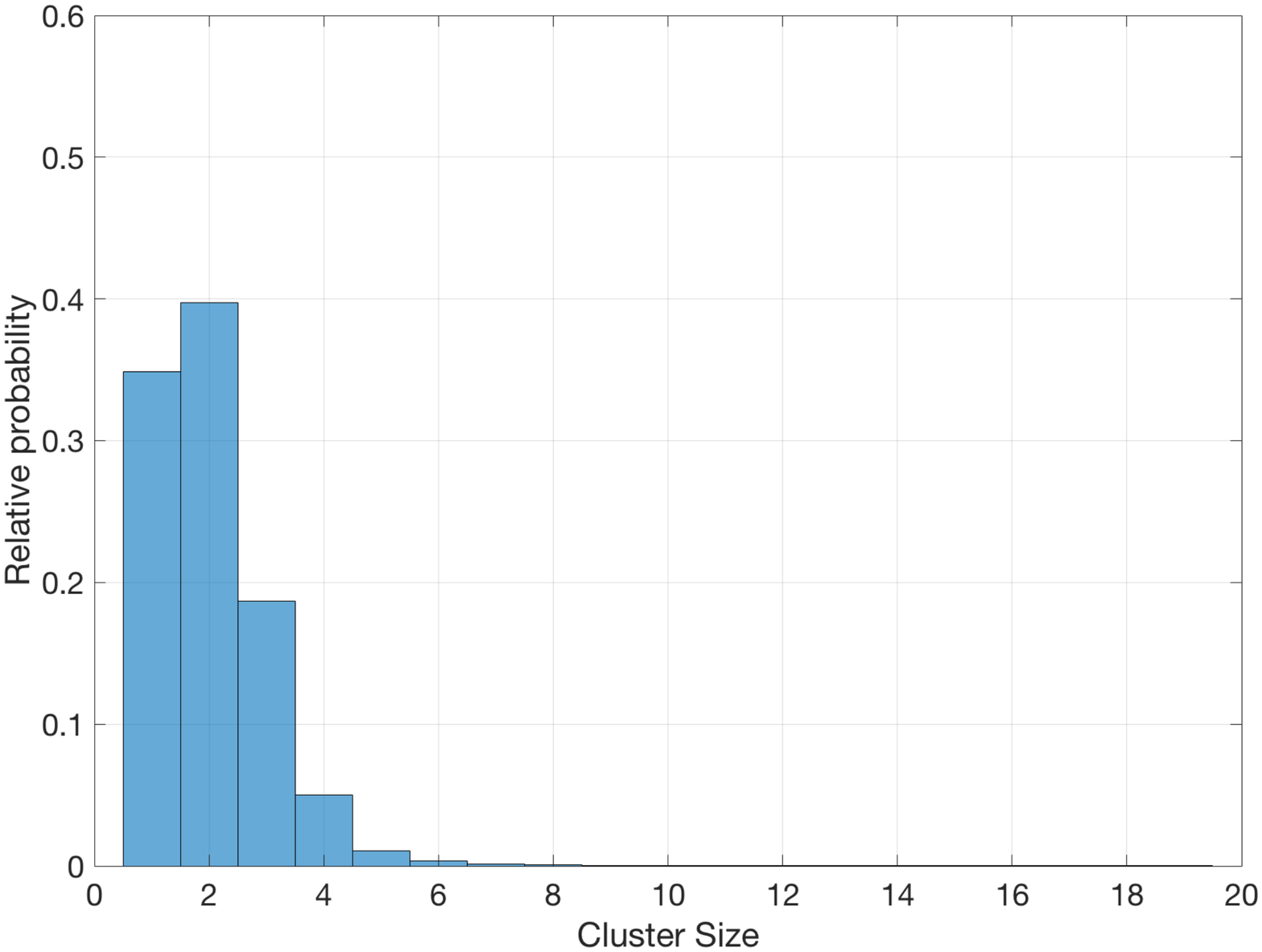}
     }
     \caption{Cluster size relative probability with the kmeans++ clustering algorithm. Setup: $\rho = 10^{-2}$ users/km$^2$ and $K_{avg}=2$.}
     \label{fig:ClustSize}
\end{figure}

The metric on which the clustering procedure is performed also impacts the performance. To highlight this dependency, let us denote by $\overline{\eta}_{chann}$ and $\overline{\eta}_{dist}$ the average rate obtained with the channel coefficients and Euclidean distance similarities, respectively, where for the sake of clarity we also neglected the dependency on $K$ and $\rho$. Figure~\ref{fig:RateGain} shows the gain in average spectral efficiency of the channel coefficients similarity with respect to the Euclidean distance similarity, \emph{i.e.}, $\overline{\eta}_{chann}-\overline{\eta}_{dist}$, for the considered clustering algorithms\footnote{We did not include the UpperBound algorithm and focused on the others for the sake of clarity in the figures.} and user densities as a function of the cluster size. It can be noticed that, for both the Random and MaxDist algorithms, the channel coefficients similarity always provides larger average rates and this gain is more evident at lower user densities and larger cluster sizes. This is due to the fact that, with the Euclidean similarity metric, a large cluster size in a low density environment implies that clusters are built with users that are located far away from each other, which easily results in significantly different channel conditions, leading to poorly representative precoding vectors. With the channel coefficient similarity, even though users are sparsely located in the Euclidean space, similar channel vectors can be found. When increasing the user densities, the difference between the two approaches decreases, since, in the Euclidean space, closer users are more likely to have similar channel conditions. Focusing on the kmeans++ algorithm, it can be noticed that the impact of the similarity metric is different. In particular, at low cluster sizes the Euclidean distance similarity provides larges average rate and this behaviour is more evident at large user densities; for low user densities and large cluster sizes, the channel coefficients similarity is the best metric as for the other algorithms. This is due to a problem known in clustering theory as the \emph{curse of dimensionality}, \cite{Curse1}: when the dimensionality of the similarity metric space is increased, as from the 2D Euclidean space to the 2$N_B$-dimensional channel coefficient space, the data points become increasingly sparse. Since with the kmeans++ algorithm we have a variable-size partitioning, the consequence is that: i) in low density volumes of the similarity space, users will be grouped in small, eventually single-element, clusters; and ii) in densely populated volumes, clusters will be much larger. Clearly, the average cluster size will still remain the one fixed \emph{a priori}, but in the channel coefficient space the consequence of the increased dimensionality is that: i) more users will be served in unicast; and ii) more clusters will be larger with respect to the 2D Euclidean space, as shown in Figure~\ref{fig:ClustSize}. In particular, the Euclidean space similarity provides clusters with a maximum of $7$ users, while with the channel coefficient similarity this can be as large as $19$. Since, in general, the average rate is larger at low cluster sizes, the average rate with the channel coefficients similarity tends to be worse, even though slightly, \emph{i.e.}, a $0.02$ bit/s/Hz loss at the most, than the Euclidean distance. The dimensionality issue has a limited impact in the MaxDist and Random algorithms thanks to the fixed-size clustering approach.

\begin{figure}[!t]
     \subfloat[$\rho=1.25\cdot 10^{-3}$ users/km$^2$.\label{fig:CDF_stdSINR_PAC_CHANN_L13}]{%
       \includegraphics[width=0.5\textwidth]{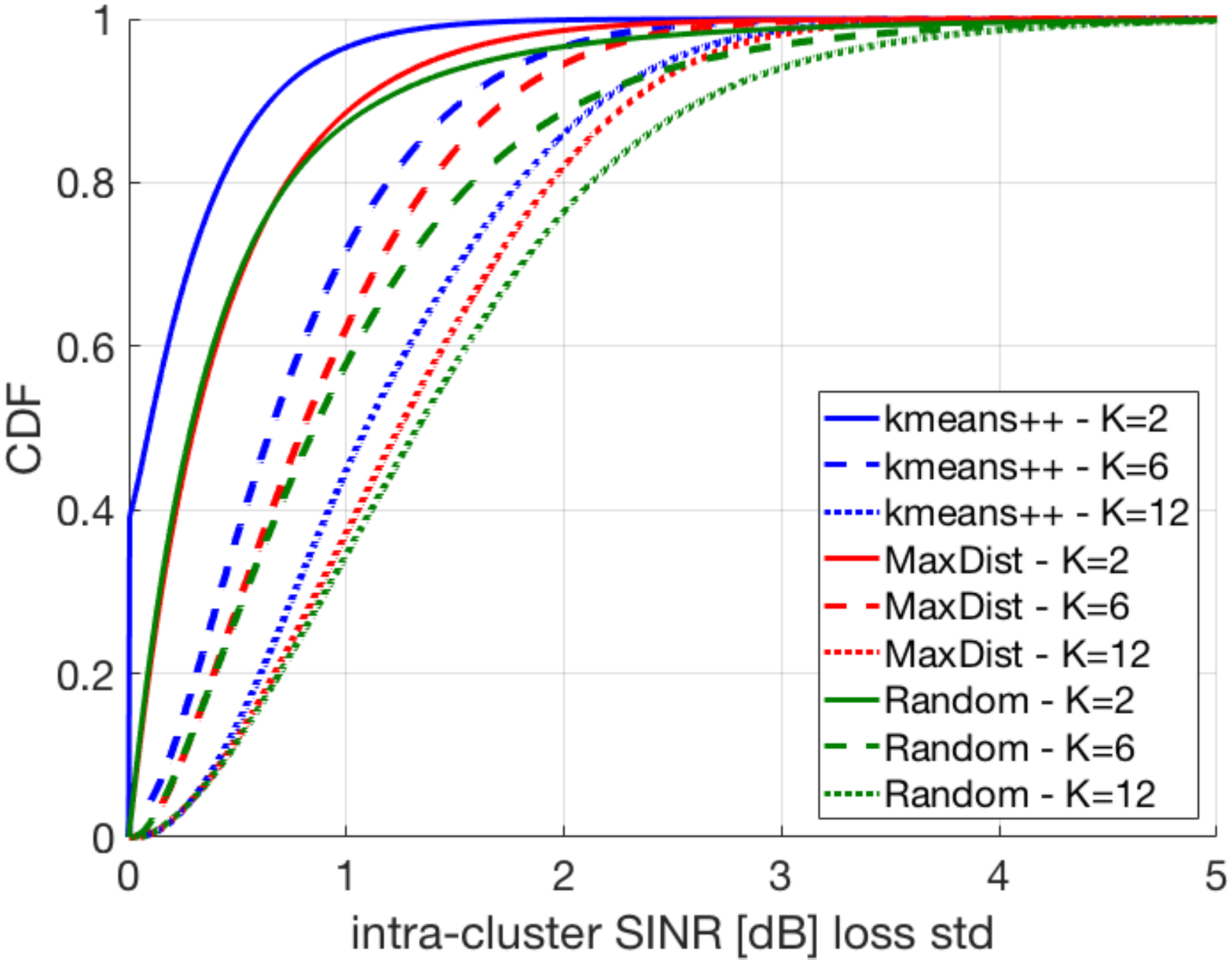}
     }
     \hfill
     \subfloat[$\rho=2.5\cdot 10^{-3}$ users/km$^2$.\label{fig:CDF_stdSINR_PAC_CHANN_L25}]{%
       \includegraphics[width=0.5\textwidth]{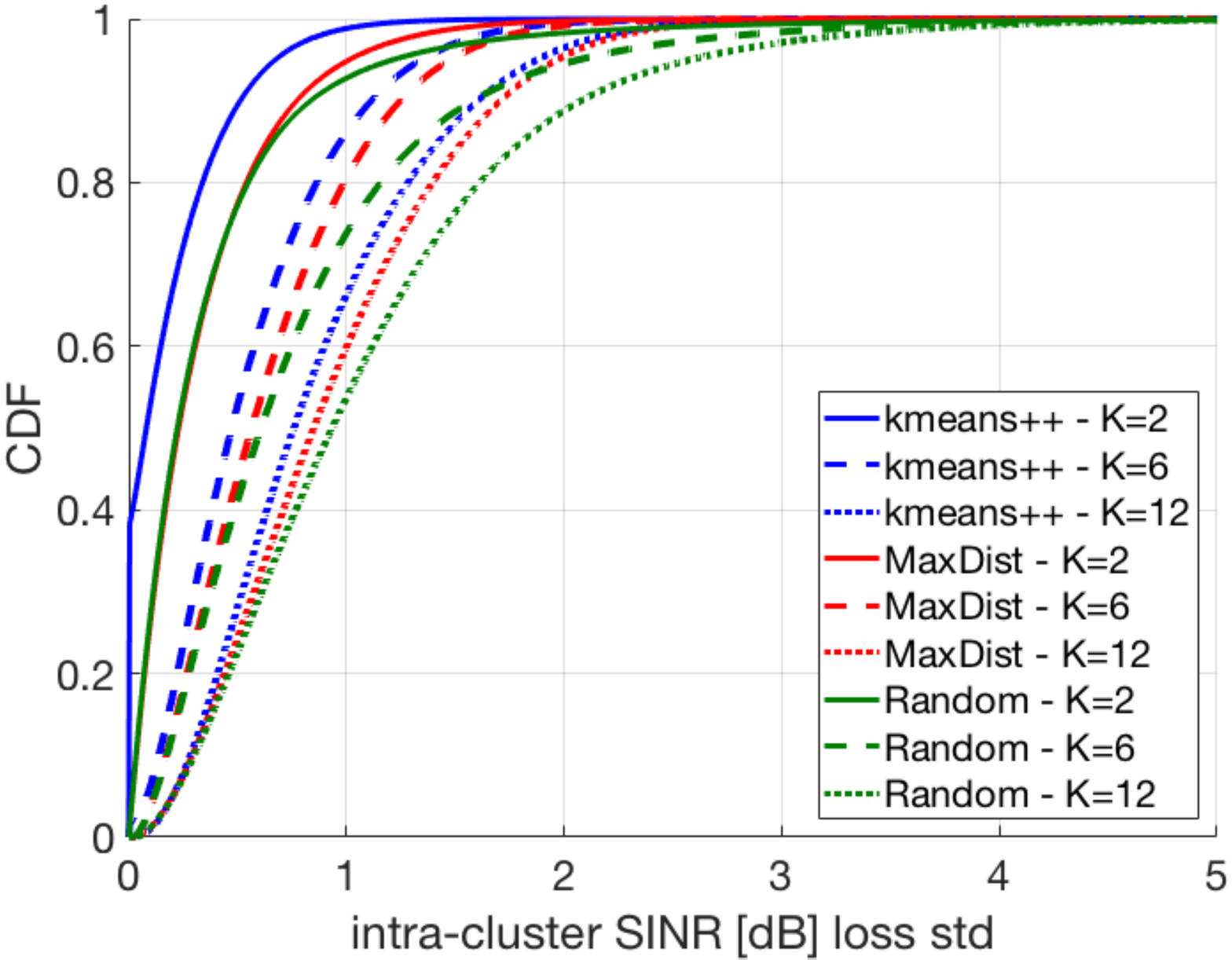}
     }
     \caption{Cumulative Distribution Function of the intra-cluster SINR loss $\Delta\gamma_{c.j}^{(b)}$ standard deviation with MMSE PAC precoding and channel coefficients similarity.}
     \label{fig:stdSINR}
\end{figure}

The better performance with increasing user densities is not only related to the larger average rate, but also to the SINR loss that users who might use a better ModCod experience due to the multicasting approach. Since the user with the lowest SINR is the one driving the serving rate, as discussed in Section~\ref{sec:SystemModel}, when the precoding vector is more representative of the actual channels that each user experiences, \emph{i.e.}, lower cluster size or larger user densities, the performance is improved also in the limitation of such loss. Let us denote the SINR of the generic $i$-th user within the $c$-th cluster of the $b$-th beam as $\gamma^{(b)}_{c,i}$ and the related cluster serving SINR by $\widetilde{\gamma}_c^{(b)} = \min_{i\in\mathcal{C}_c^{(b)}}\left\{\gamma^{(b)}_{c,i}\right\}$. The performance loss that the users in the cluster experience due to average precoding is given by $\Delta\gamma_{c.i}^{(b)} = \gamma^{(b)}_{c,i}-\widetilde{\gamma}_c^{(b)}$ and it is null for the worst-case user only, \emph{i.e.}, $\Delta\gamma_{c.j}^{(b)}=0$ iff $\gamma^{(b)}_{c,j}=\widetilde{\gamma}_c^{(b)}$, $j\in\mathcal{C}_c^{(b)}$. Figure~\ref{fig:stdSINR}, the Cumulative Distribution Function of the standard deviation of $\Delta\gamma_{c.j}^{(b)}$, $\sigma_{\Delta\gamma}$, computed for all beams, clusters, and users, is shown for different user densities and cluster sizes for the considered clustering algorithms. It can be noticed that, as expected, by increasing the cluster size the standard deviation $\sigma_{\Delta\gamma}$ moves to the right, \emph{i.e.}, to worst scenarios in which more users experience larger losses with respect to the minimum SINR user driving the selected ModCod. In addition, by comparing Figures \ref{fig:CDF_stdSINR_PAC_CHANN_L13} and \ref{fig:CDF_stdSINR_PAC_CHANN_L25}, it can be noticed that an increased user density limits the SINR loss due to the larger cluster sizes, in particular for lower cluster sizes. Since the same behaviour can be noticed with both PAC and SPC precoding, in the following we focus on the performance with PAC MMSE precoding for the sake of clarity.

\begin{figure}[!t]
    \centering
    \includegraphics[width=0.5\columnwidth]{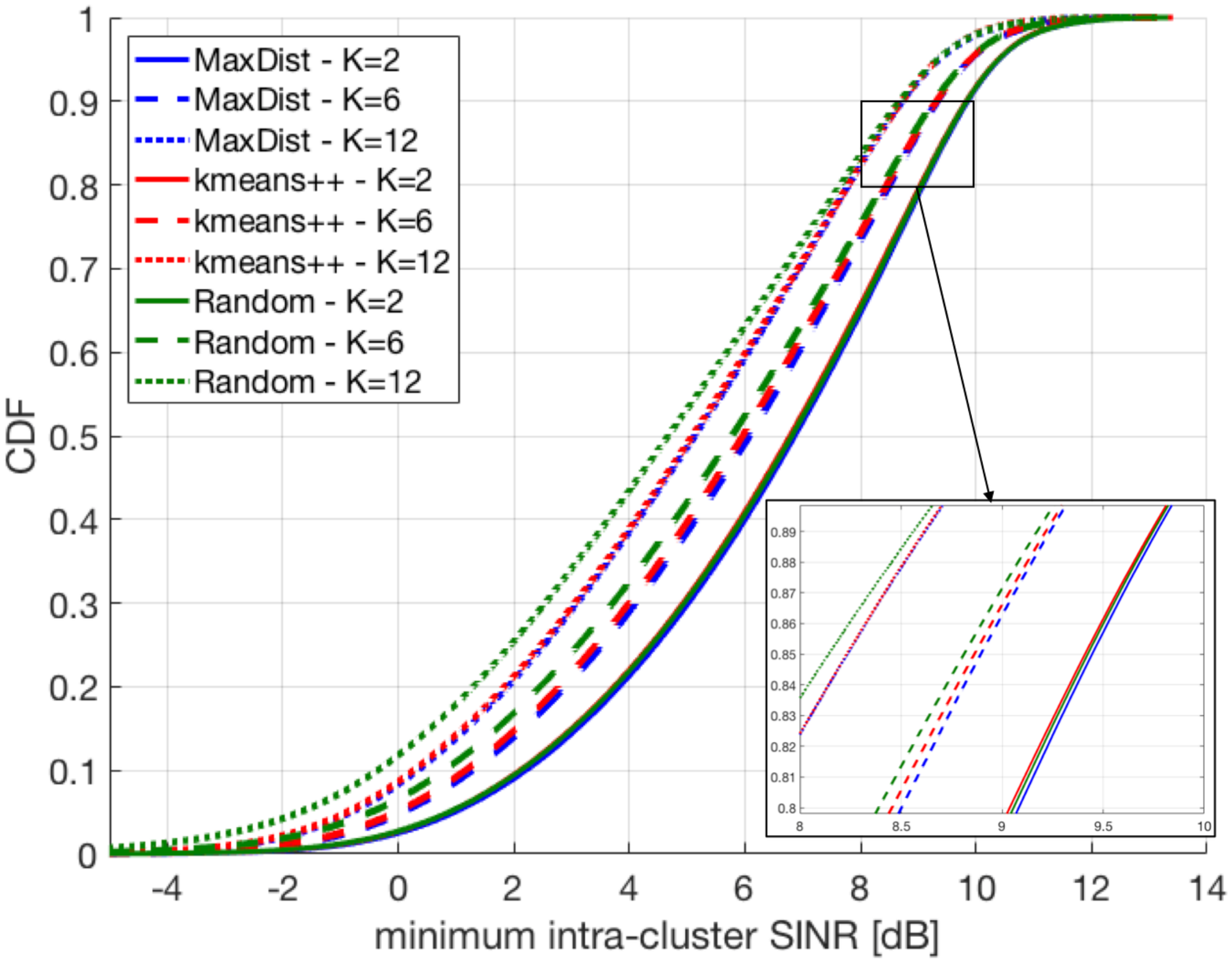}
    \caption{Cumulative Distribution Function of the minimum intra-cluster SINR $\widetilde{\gamma}_c^{(b)}$ with MMSE PAC precoding. Setup: $\rho = 2.5\cdot 10^{-3}$ users/km$^2$.}
    \label{fig:CDF_minSINR_PAC_CHANN_L25}
\end{figure}

\subsection{Clustering Algorithms Performance}
Based on the performance in terms of average rate as a function of the cluster size shown in Figures~\ref{fig:Rate_PAC_L}-\ref{fig:Rate_SPC_H}, it can be noticed that the UpperBound algorithm indeed upper bounds the performance of the other solutions in all the considered cases. As discussed in the previous sections, this is due to the fact that this algorithm does not take into account the system outliers, which is one of the most limiting factor in the context of clustering algorithms.\\
With respect to the Random algorithm, both the MaxDist and kmeans++ algorithms proposed in this work provide better performance almost independently from the cluster size, user density, and precoding algorithm. The only exception to this is when kmeans++ is implemented with large user densities and low to medium cluster sizes, in which case the performance is below that provided by the Random algorithm when the channel coefficients similarity is the chosen metric, as in Figures \ref{fig:CAPACITY_CHANN_PAC_H} and \ref{fig:CAPACITY_CHANN_SPC_H}. The reason for this behaviour of the kmeans++ algorithm has been discussed in the previous section. At low user densities, the kmeans++ algorithm outperforms the MaxDist solution. As a matter of facts, a variable-size clustering algorithm is more suitable to environments in which users are particularly spread in the considered space, \emph{i.e.}, when the majority of the users can be considered as outliers. In fact, in this case, the variable-size clustering algorithm will form unicast clusters for users that are far away from all the others, while grouping together the few users that are close to each other, thus strongly limiting the impact of the outliers. A fixed-size clustering algorithm is forced to group together $K$ users in any case and, thus, it greatly suffers from the outliers problem. When the user density is increased, the number of outlier users significantly decreases and this has two consequences: i) the fixed-size algorithms (MaxDist and Random) will be able to always find users that have similar channel conditions and, thus, will provide good performance; ii) the kmeans++ algorithm will suffer from the variable-size principle, since, although being a minimum partition clustering algorithm that minimises the cost function in eq.~(\ref{eq:SSEcost}) in the considered metric space, the presence of clusters with cardinality larger than $K$ limits the performance as the channel coefficients vectors used to build the average precoding vector will be less representative for the cluster users. This aspect is highlighted in Figures \ref{fig:CDF_stdSINR_PAC_CHANN_L25} and \ref{fig:CDF_minSINR_PAC_CHANN_L25}. In particular, while the CDF of the intra-cluster SINR loss $\Delta\gamma_{c.j}^{(b)}$ standard deviation in Figure~\ref{fig:stdSINR} is significantly lower for the kmeans++ algorithm, substantiating its minimum variance property, the CDF of the clusters' minimum SINR $\widetilde{\gamma}_{c}^{(b)}$ in Figure~\ref{fig:CDF_minSINR_PAC_CHANN_L25} is worse with low cluster size ($K=2$), and almost identical to the MaxDist and better than the Random algorithms with medium and large cluster sizes ($K=6$ and $K=12$, respectively). Thus, even if the variance in the similarity space is indeed minimised, the average channel vectors are less representative and the minimum SINR is lower.

\begin{figure}[!t]
     \subfloat[$\rho=1.25\cdot 10^{-3}$ users/km$^2$.\label{fig:RATEvsPOW_CHANN_PAC_L}]{%
       \includegraphics[width=0.5\textwidth]{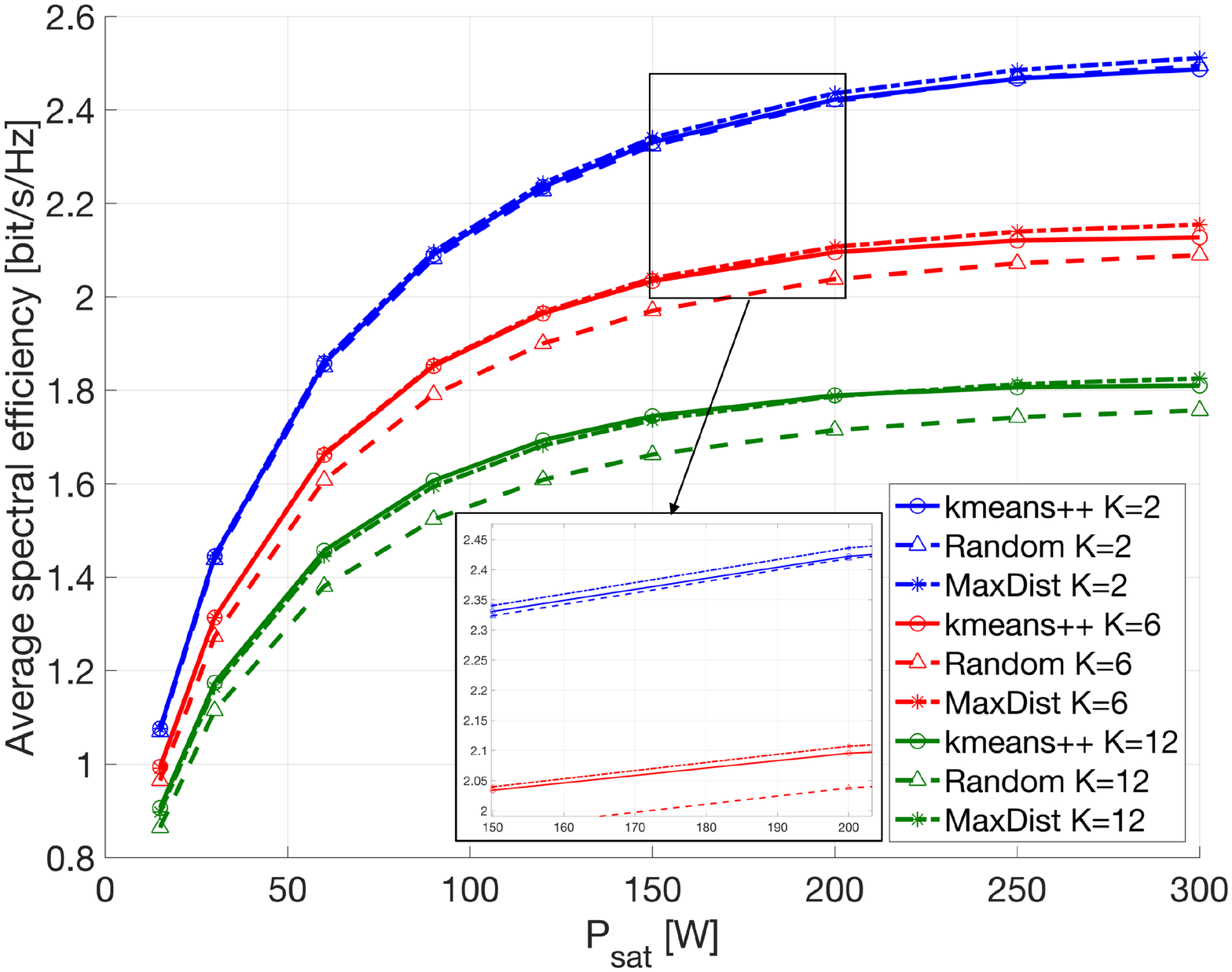}
     }
     \hfill
     \subfloat[$\rho=0.01$ users/km$^2$.\label{fig:RATEvsPOW_CHANN_PAC_H}]{%
       \includegraphics[width=0.5\textwidth]{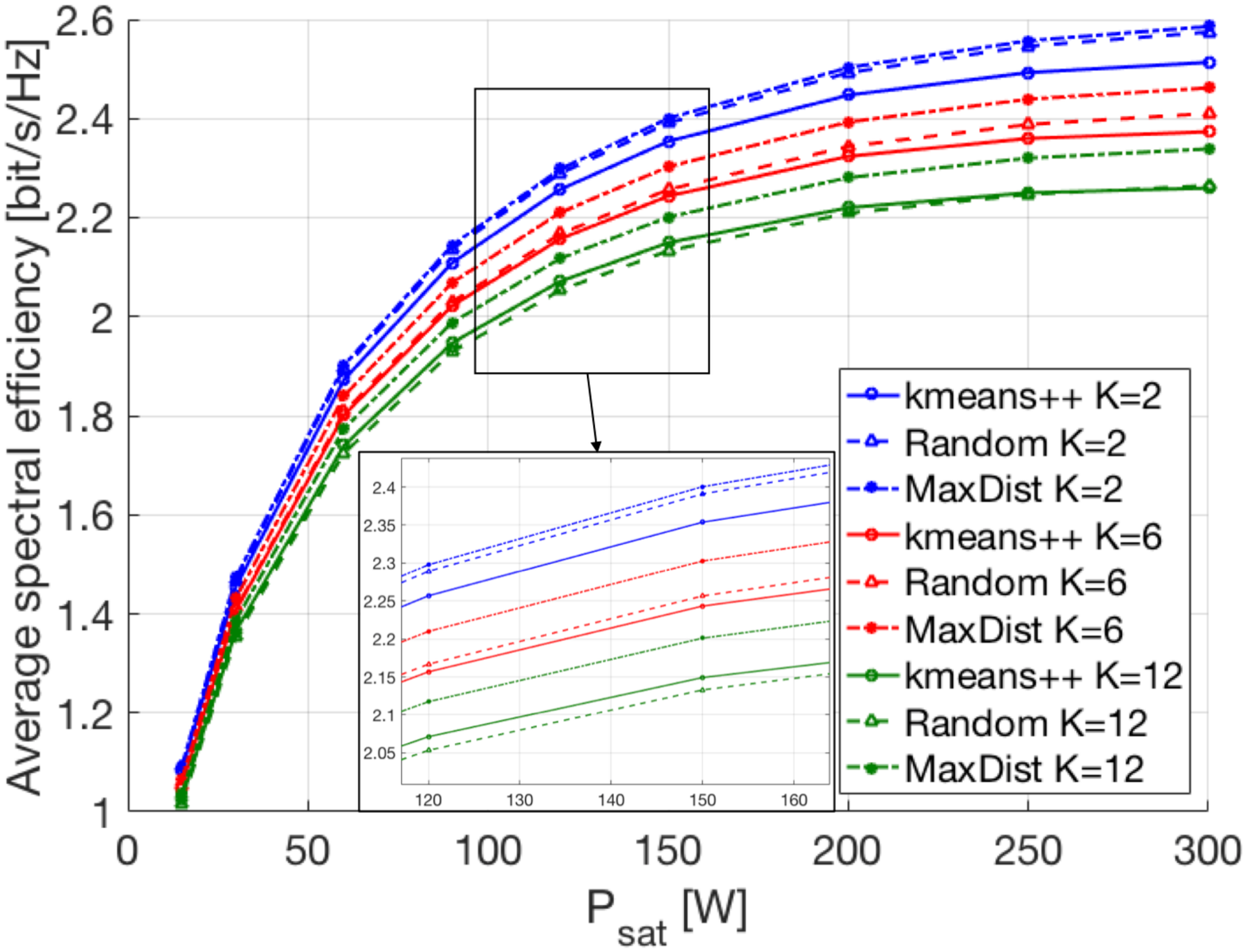}
     }
     \caption{Average rate as a function of $P_{sat}$ with MMSE PAC precoding and channel coefficients similarity.}
     \label{fig:RATEvsPOW_CHANN_PAC}
\end{figure}

Finally, Figure~\ref{fig:RATEvsPOW_CHANN_PAC} shows the average rate as a function of the transmitted power $P_{sat}$ for different cluster sizes. These figures show that the above observations on the performance of the different clustering algorithms hold even for different transmitted power levels. In particular, apart from a limited transition range at very low power levels, namely between $15$ and $50$ Watts, the average rate has the same behaviour in all algorithms for increasing $P_{sat}$ up to the saturation point in which the precoder cannot yield any further improvement due to the severe interference.

%\begin{figure}[!t]
%     \subfloat[Euclidean distance.\label{fig:CDF_minSINR_PAC_DIST_L25}]{%
%       \includegraphics[width=0.5\textwidth]{CDF_minSINR_PAC_DIST_L25_v3.png}
%     }
%     \hfill
%     \subfloat[Channel coefficients.\label{fig:CDF_minSINR_PAC_CHANN_L25}]{%
%       \includegraphics[width=0.5\textwidth]{CDF_minSINR_PAC_CHANN_L25_v2.png}
%     }
%     \caption{Cumulative Distribution Function of the minimum cluster SINR $\widetilde{\gamma}_c^{(b)}$ with MMSE PAC precoding. Setup: $\rho = 2.5\cdot 10^{-3}$ users/km$^2$.}
%     \label{fig:minSINR}
%\end{figure}

\subsection{Computational complexity}
To conclude the overall analysis of the considered clustering algorithms, we now provide the time complexity analysis. To this aim, we focus on a single iteration of the considered algorithms, \emph{i.e.}, the complexity for a single beam clusterisation, and we also assume to be in a worst-case scenario in which $N_K^{(b)} = N_K = \max_b\left\{N_K^{(b)}\right\}$ and $N_U^{(b)} = N_U = \max_b\left\{N_U^{(b)}\right\}$. With respect to the k-means++ algorithm, its computational complexity is known from the literature and it is given by $\mathcal{O}\left(\log N_K\right) = \mathcal{O}\left(\log\frac{N_U}{K}\right) \sim \mathcal{O}\left(\log N_U\right)$, since $N_K = N_U/K$.\\
The computational complexity of the Random clustering algorithm can be computed by taking into account the following single-step complexities: i) random user selection (step 4): $\mathcal{O}\left(N_U\right)$ for accessing the user indexes array and $\mathcal{O}\left(1\right)$ for the random selection of the reference user; ii) $d_{qj} = \left.\parallel\mathbf{u}_{q}^{(b)}-\mathbf{u}_{j}^{(b)}\parallel\right.$, $\forall j\in \mathcal{Q}^{(b)}$ (step 5): we have $2\mathcal{O}\left(d\right)$, since both the $d$-dimensional subtraction and summation are $\mathcal{O}\left(d\right)$, plus $\mathcal{O}\left(d\right)$ for the squares and $\mathcal{O}(1)$ for the square root. These operations are to be repeated for $\left|\mathcal{Q}^{(b)}\right|$ times, which leads to $\mathcal{O}\left(d\left|\mathcal{Q}^{(b)}\right|\right)\sim\mathcal{O}\left(\left|\mathcal{Q}^{(b)}\right|\right)$ that depends on the specific iteration number; iii) sort algorithm to find $K$ minimum values in a $\left|\mathcal{Q}^{(b)}\right|$ elements array (step 6): $\mathcal{O}\left(\left|\mathcal{Q}^{(b)}\right|^2\right)$; and iv) update $\mathcal{Q}^{(b)}$ (step 7): $\mathcal{O}\left(\left|\mathcal{Q}^{(b)}\right|\right)$ to access the array. Thus, noting that if we denote by $\ell$ the step in which the clustering algorithm is operating we have $\left|\mathcal{Q}^{(b)}\right| = N_U-\ell N_K$, we can compute the overall Random clustering time complexity, for each beam, as follows:
\begin{eqnarray*}
    f^{(rand)}\left(N_U\right) &\sim & \sum_{\ell=1}^{N_K} \left[ \mathcal{O}\left(1\right) + \mathcal{O}\left(N_U-\ell N_K\right) \right.\\
    && \left. + \mathcal{O}(d)\mathcal{O}\left(N_U-\ell N_K\right) +\right. \\
    && + \left.\mathcal{O}\left({\left(N_U - \ell K\right)}^2\right) + \mathcal{O}\left(N_U-\ell N_K\right) \right]\\
    &\overset{(a)}{\sim}& \sum_{\ell=1}^{N_K} \left[ \mathcal{O}\left(N_U-\ell N_K\right) + \mathcal{O}\left({\left(N_U - \ell K\right)}^2\right)\right]\\
    &\overset{(b)}{\sim}& \mathcal{O}\left(N_U^2\right)
\end{eqnarray*}
where: i) in $(a)$ we exploited the fact that $\mathcal{O}\left(1\right) + \mathcal{O}\left(N_U-\ell N_K\right)\sim \mathcal{O}\left(N_U-\ell N_K\right)$ and $\mathcal{O}(d)\mathcal{O}\left(N_U-\ell N_K\right)\sim \mathcal{O}\left(N_U-\ell N_K\right)$; and ii) in $(b)$ we noticed that $\mathcal{O}\left({\left(N_U - \ell K\right)}^2\right)$ is the dominant term and then expanded the square. It can be easily shown that the time complexity of the Upper Bound algorithm is the same as that for the Random one, since in both cases the dominant operation is the sorting procedure.\\
Finally, with respect to the MaxDist algorithm, the only difference with respect to the Random clustering algorithm is that, at each iteration, the barycentre of the not yet clustered users shall be computed and its distance from all of the remaining users sorted. These operations have a time complexity of $\mathcal{O}(d)\mathcal{O}\left(N_U-\ell N_K\right)$ and $\mathcal{O}\left({\left(N_U - \ell K\right)}^2\right)$, respectively. Once we include these two additional terms in the above equations, it is straightforward to note that the overall complexity is still $\mathcal{O}\left(N_U^2\right)$. In conclusion, the computational complexity of the MaxDist algorithm is the same as the Random and Upper Bound algorithms, $\mathcal{O}\left(N_U^2\right)$, while the kmeans++ algorithm has a $\mathcal{O}\left(\log N_U\right)$ complexity. Thus, the kmeans++ algorithm has a significantly lower time complexity with respect to the MaxDist, Random, and Upper Bound clustering approaches.

\section{Conclusions}
\label{sec:Conclusion}
In this paper, we focused on users clustering for multicast precoding in multi-beam satellite systems. In particular, moving from early works, we defined a mathematical framework for the design of multicast precoding systems based on clustering theory and algorithms. In this context, two clustering algorithms have been proposed: i) a fixed-size clustering, aimed at limiting the impact of outlier users; and ii) a variable-size clustering based on the kmeans++ algorithm. The performance of the proposed algorithms has been compared to existing algorithms for both PAC and SPC MMSE precoding and for variable user densities. Two similarity metrics have been considered for clustering: i) the Euclidean distance; and ii) the distance in the multi-dimensional space of user channel coefficients. Numerical simulations showed that the achievable rate obtained with the proposed fixed-size clustering algorithm always outperforms the solutions available in the literature, almost reaching the upper bound performance. As for the variable-size clustering algorithm, the performance is better than already available solutions; with increasing user densities and channel coefficient similarity fixed-size solutions shall be preferred. The algorithms have also been compared in terms of achievable rate as a function of the satellite transmitted power. In addition, the clustering performance has been assessed also in terms of the representativeness of the average precoding vector with respect to the actual user channel vectors. Finally, the computational complexity of the proposed algorithms has also been assessed. In general, the following conclusions hold: i) for fixed-size clustering, the channel coefficients similarity metric provides better performance with respect to the Euclidean distance; ii) variable-size clustering provides slightly worse performance with respect to fixed-size solutions. Future developments of this paper will take into account, among the others, the inclusion of non-uniform user traffic requests and scheduling aspects.

\section*{Acknowledgment}
This work has been partially supported by European Space Agency (ESA) funded project OPTIMUS (``OPtimized transmission TechnIques for satcoM UnicaSt Interactive traffic''), contract 4000116421/15/NL/FE, and SatNEx IV COO2-PART1 WI4 ``Forward Packet Scheduling Strategies for Emerging Satellite Broadband Networks.'' The views of the authors of this paper do not reflect the views of ESA.

\ifCLASSOPTIONcaptionsoff
  \newpage
\fi

\end{document}